\documentclass[conference]{IEEEtran}

\usepackage{epsf}
\usepackage{graphicx,color}
\usepackage{amsmath,bm}
\usepackage{amssymb}
\usepackage{epsfig,latexsym,amsmath,epsf,amssymb,amsfonts}
\usepackage{verbatim}
\usepackage{placeins}
\usepackage[hang]{subfigure}
\usepackage{epstopdf}
\usepackage{cite}
\usepackage{amsthm}
\usepackage{graphics}

\newtheorem{theorem}{Theorem}
\newtheorem{lemma}{Lemma}
\newtheorem{corollary}{Corollary}

\newcommand{\ignore}[1]{}

\usepackage{url}

\usepackage{algorithm}
\usepackage{algorithmic}

\usepackage[capitalise]{cleveref}

\usepackage{filecontents}

\begin{document}
\title{Asymmetric Degrees of Freedom of the Full-Duplex MIMO 3-Way Channel with Unicast and Broadcast Messages}
\author{
	\large Adel M. Elmahdy$^{\dag\S}$, Amr El-Keyi$^{\dag}$, Yahya Mohasseb$^{\star}$, Tamer ElBatt$^{\dag\P}$,\\ 
		   Mohammed Nafie$^{\dag\P}$, Karim G. Seddik$^{\ddag}$, and Tamer Khattab$^{\S}$\\ [.1in]
	\normalsize
	\begin{tabular}{c}
		$^{\dag}$Wireless Intelligent Networks Center (WINC), Nile University, Giza, Egypt.\\
		$^{\star}$Dept. of Communications, The Military Technical College, Cairo, Egypt.\\
		$^{\P}$Dept. of EECE, Faculty of Engineering, Cairo University, Giza, Egypt.\\
		$^{\ddag}$Electronics and Communications Engineering Dept., American University in Cairo, New Cairo, Egypt.\\
		$^{\S}$Electrical Engineering Dept., Qatar University, Doha, Qatar.\\
		adel.elmahdy@ieee.org, aelkeyi@nu.edu.eg, \{mohasseb, telbatt, mnafie\}@ieee.org,
		kseddik@aucegypt.edu, tkhattab@ieee.org
	\end{tabular}
}
\maketitle
%
\makeatletter{\renewcommand*{\@makefnmark}{}
\footnotetext{\hrule \vspace{0.05in}
The research work of A. M. Elmahdy and T. ElBatt was made possible by grants number NPRP 4-1034-2-385 and NPRP 5-782-2-322 from the Qatar National Research Fund, QNRF (a member of Qatar Foundation, QF). 
The research work of A. El-Keyi, M. Nafie and T. Khattab was made possible by grant number NPRP 7-923-2-344 from the QNRF. 
The statements made herein are solely the responsibility of the authors.
%
}\makeatother}


\begin{abstract}
In this paper, we characterize the asymmetric total degrees of freedom (DoF) of a multiple-input multiple-output (MIMO) 
3-way channel. 
Each node has a separate-antenna full-duplex MIMO transceiver with a different number of antennas, where each antenna can 
be configured for either signal transmission or reception. 
We study this system under two message configurations; the first configuration is when each node has two unicast messages 
to be delivered to the two other nodes, while the second configuration is when each node has two unicast messages as well 
as one broadcast message to be delivered to the two other nodes. 
For each configuration, 
we first derive upper bounds on the total DoF of the system. Cut-set bounds in conjunction with 
genie-aided bounds are derived to characterize the achievable total DoF. 
Afterwards, we analytically derive the optimal number of transmit and receive antennas at each node to maximize the total 
DoF of the system, subject to the total number of antennas at each node.
Finally, the achievable schemes for each configuration are constructed. 
The proposed schemes are mainly based on zero-forcing and null-space transmit beamforming.
\end{abstract}


\section{Introduction} \label{sec:intro}
Interference-limited wireless communication networks have been extensively investigated over recent years. 
Despite the fact that uncoordinated interference decreases the achievable data rates in wireless networks, novel paradigms have emerged to sagaciously harness interference and, hence, efficiently utilize the scarce spectrum and enhance the network performance.

Full-duplex systems have attracted a great deal of attention recently due to their potential benefits to significantly enhance the throughput and spectral efficiency of conventional half-duplex systems \cite{IBFD_Survey}.
Existing wireless communication systems operate in either a time-division duplex or a frequency-division duplex mode to separate the downlink and uplink traffic.
However, recent results from academia \cite{Rice_FullDuplex,MIDU,Exp_FullDuplex,FullDyplex_WiFi,Passive_SelfInterf_FD,SingAnt_FD,Katti_fullDuplexRadios,5G_FullDuplex} and industry \cite{KumuNetworks} have proposed various practical designs to implement in-band full-duplex radios by cancelling or suppressing the self-interference signal, generated during simultaneous transmission and reception, at the RF and baseband level.
There are two possible methods of antenna interfacing for full-duplex MIMO transceivers; 
separate-antenna architecture \cite{Rice_FullDuplex,MIDU,Exp_FullDuplex,FullDyplex_WiFi,Passive_SelfInterf_FD}, and 
shared-antenna architecture \cite{SingAnt_FD,Katti_fullDuplexRadios,5G_FullDuplex}.
In separate-antenna architecture, each antenna is dedicated to either signal transmission or reception. 
In shared-antenna architecture, each antenna simultaneously transmits and receives signals on the same channel with the aid of a circulator that routes the transmitted signal from the TX signal chain to the antenna and the received signal on the antenna to the RX signal chain.
Full-duplex systems are envisioned to have an enormous impact on the evolution of future 5G generations of wireless communication systems.

The two-way communication channel was introduced in the seminal paper by Shannon \cite{Shannon_2way}.
The extension of the two-way channel to the case of three nodes, i.e., the 3-way channel, has recently attracted much attention
\cite{D2D_Chabban,3way_CapReg_Chabban,3way_Channel_Chabaan}.
It is assumed that all nodes operate in a perfect full-duplex mode.
Furthermore, there are six unicast messages to be exchanged among the nodes; each node is intended to exchange unicast messages with the other nodes simultaneously.
The sum-capacity of the 3-way channel, that characterizes the DoF of the channel, is studied in \cite{D2D_Chabban} for the Gaussian channel model.
It is shown that the sum-capacity is achievable within a gap of 2 bits.
The achievable transmission strategy is to allow the two nodes with the strongest channel coefficient to communicate while leaving the third node silent.

On the other hand, the capacity region of the 3-way channel is considered for the linear shift deterministic channel model with reciprocal channel gains in \cite{3way_CapReg_Chabban}.
Under this framework, the outer bounds of the 3-way channel 
are related to those of the linear shift deterministic Y-channel \cite{Det_YChannel_Chabaan} through $\Delta \text{-} \text{Y}$ transformation, inspired from electrical circuit theory.
The capacity achieving schemes are mainly based on multi-way relaying by signal alignment, interference neutralization and backward decoding.

The authors in \cite{3way_Channel_Chabaan} investigate the symmetric DoF of a MIMO 3-way channel with homogeneous antenna configurations; each node has $M_T$ transmit antennas and $M_R$ receive antennas.
Cut-set bounds and genie-aided upper bounds are derived to characterize the symmetric total DoF of the channel.
Then, the authors propose achievable schemes for the derived total DoF based on null-space beamforming and MIMO interference alignment.

\subsection{Summary of Results}
The main contribution of this paper is the characterization of the total DoF of a MIMO 3-way channel with heterogeneous antenna configurations.
Each node has a separate-antenna full-duplex MIMO transceiver where each antenna can be configured to either transmit or receive, and the nodes have different numbers of antennas.
In particular, node $\ell$, where $\ell \!\in\! \left\{1,2,3\right\}$, has a total of $M_{\ell}$ antennas with $M_{T_{\ell}}$ antennas utilized 
for signal transmission and $M_{R_{\ell}} \!=\! M_{\ell} - M_{T_{\ell}}$ antennas used for signal reception.
Moreover, without loss of generality, the total number of antennas at each node is in such a way that 
$M_1 \!\geq\! M_2 \!\geq\! M_3$ for nodes 1, 2 and 3, respectively.
It should be noted that the proposed system model is a generalized version of the symmetric model studied by Maier~\emph{et~al.} in \cite{3way_Channel_Chabaan} where the total number of antennas of each node are the same, and each node has $M_T$ transmit antennas and $M_R$ receive antennas.
Furthermore, we study this system under two message configurations; first, each node has two unicast messages to be delivered to the two other nodes, and, second, each node has two unicast messages as well as one broadcast message to be delivered to the two other nodes.

For each message configuration, we first derive upper bounds on the total DoF of the system in terms of $M_{T_{\ell}}$ and $M_{R_{\ell}}$, where $\ell \!\in\! \left\{1,2,3\right\}$. Under the unicast message configuration, cut-set bounds in conjunction with genie aided bounds are utilized to characterize the achievable total DoF in this case. 
On the other hand, under the unicast and broadcast message configuration, the cut-set bounds are achievable. 
It~should be noted that a broadcast message is considered as a desired message by all nodes and it is not treated as interference.
Therefore, unlike the unicast message configuration, a broadcast message gives an additional degree of freedom and, hence, the cut-set bounds can be achieved in this case.
Afterwards, we analytically derive the optimal number of transmit and receive antennas at each node to maximize the total DoF of the system, subject to the total number of antennas at each node.
Finally, the achievable schemes for each configuration are constructed. The schemes are mainly based on zero-forcing and null-space beamforming.

\subsection{Paper Organization}
The remainder of this paper is organized as follows.
The system model and underlying assumptions are presented in Section~\ref{sec:sys_mod}.
Next, the upper bounds on the total DoF of the system, the optimal antenna allocation at each node, and the achievable schemes are derived in Section~\ref{sec:uni_only} when the system only features unicast messages,
whereas they are derived in Section~\ref{sec:uni_brcst} when the system features unicast as well as broadcast messages.
Finally, the paper is concluded in Section~\ref{sec:con}.

\subsection{Notation}
Lower and upper boldface letters are used to denote column vectors and matrices, respectively.
$\mathbf{X}^T$, $\mathbf{X}^H$ and $\mathbf{X}^{\dag}$ denote the transpose, the Hermitian transpose and the pseudo-inverse of $\mathbf{X}$, respectively.
$\mathbf{I}_{m}$ is an $m \times m$ identity matrix, and $\mathbf{0}_{m \times n}$ is an $m \times n$ zero matrix.
The sequence $\left(\mathbf{x}(1), \mathbf{x}(2), \ldots, \mathbf{x}(N) \right)$ is denoted by $\mathbf{x}^N$.
Let $\textit{h} \left(\mathbf{x}\right)$ denote the differential entropy of a random vector $\mathbf{x}$, and 
$\textit{I} \left(\mathbf{x}; \mathbf{y}\right)$ denote the mutual information between two random vectors $\mathbf{x}$ and $\mathbf{y}$. 

\section{System Model} \label{sec:sys_mod}
\begin{figure}
    \centering
    \includegraphics[width=1\linewidth]{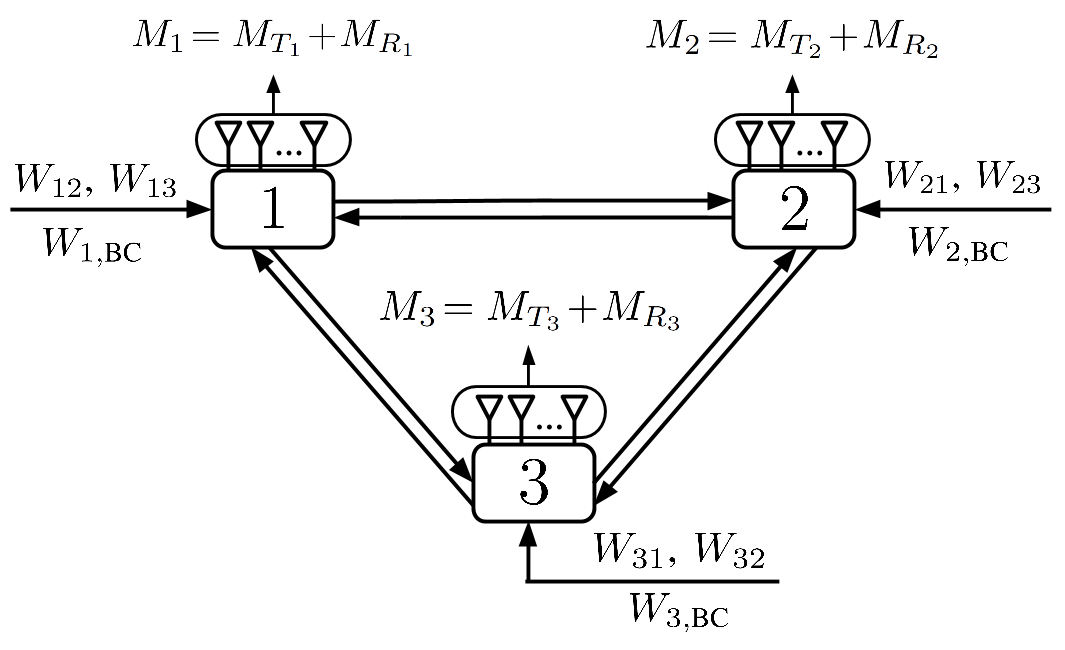}
    \caption{The system model.}
    \label{fig:sysModel}
\end{figure}

We consider the MIMO 3-node fully-connected interference network, a.k.a. the MIMO 3-way channel, depicted in Fig.~\ref{fig:sysModel}.
Each node has a separate-antenna full-duplex MIMO transceiver where each antenna can be configured for either signal transmission or reception.
Consequently, node~$\ell$, where $\ell \in \mathcal{U} \!\!=\!\! \{1,2,3\}$, has $M_{\ell}$ antennas of which it utilizes $M_{T_{\ell}}$ antennas for signal transmission and $M_{R_{\ell}}$ antennas for signal reception, where 
$M_{T_{\ell}} \!+\! M_{R_{\ell}} = M_{\ell}$.
Furthermore, our asymmetric setting entails a different number of antennas at the different nodes. 
Henceforth, without loss of generality, we assume that $M_1 \geq M_2 \geq M_3$.
Moreover, the signals as well as the channel coefficients are assumed to be complex-valued.
Similar to \cite{D2D_Chabban,3way_CapReg_Chabban,3way_Channel_Chabaan}, we assume that the nodes operate in a perfect full-duplex mode, i.e., each node can transmit and receive messages simultaneously and the effect of residual self-interference, imposed by the transmit antennas on the receive antennas within the same transceiver, is perfectly cancelled or suppressed.
It is worth mentioning that recent research results indicate that the practical implementation of separate-antenna 
in-band full-duplex MIMO transceivers is becoming technologically feasible 
\cite{Han_fullDuplx_2020,Rice_FullDuplex,MIDU,Exp_FullDuplex,FullDyplex_WiFi,Passive_SelfInterf_FD}.
%
%

The 3-way channel features two kinds of messages; unicast messages and broadcast messages.
In other words, node~$i$ can send one or more of three independent messages;
two unicast messages $W_{ij} \text{ and } W_{ik}$ to nodes $j$~and~$k$ with rates $R_{ij} \text{ and } R_{ik}$, 
respectively, and one broadcast message $W_{i,\text{BC}}$ to both node~$j$ and node~$k$ with a rate $R_{i,\text{BC}}$, 
for $i, j, k \in \mathcal{U}$~and~$i \neq j \neq k$.

The transmitted signal from node~$i$ is denoted by $\mathbf{x}_{i} \in \mathbb{C}^{M_{T_{i}} \times 1}$.
It is assumed that the power of the transmitted signal from node~$i$ is bounded by $\rho$, i.e., 
$\mathbb{E} \left\{\|\mathbf{x}_{i}\|^{2}\right\} \leq \rho$.
Taking into account the aforementioned description of the system model, the received signal at node~$j$ at time slot $n$, denoted by $\mathbf{y}_j \!\left(n\right) \in \mathbb{C}^{M_{R_{j}} \times 1}$, is given by
\begin{equation} \label{eqn:rx_sig}
	\displaystyle
    \mathbf{y}_j \!\left(n\right) = \sum_{i \in \mathcal{U}, i \neq j}
    \mathbf{H}_{ij} \: \mathbf{x}_{i} \!\left(n\right)
    + \mathbf{z}_j \!\left(n\right),
\end{equation}
where $\mathbf{H}_{ij} \in \mathbb{C}^{M_{R_{j}} \times M_{T_{i}}}$ is the random channel matrix from node~$i$ to node~$j$,
and $\mathbf{z}_j \in \mathbb{C}^{M_{R_{j}} \times 1}$ is the additive noise signal at node~$j$ whose elements are independent and identically distributed (i.i.d.) complex Gaussian random variables with zero mean and unit variance.
Throughout this paper, we assume that each node has perfect knowledge of the channel state information (CSI) from the other two nodes. 
Moreover, for the sake of notational simplicity, we drop the time index~$n$ throughout the sequel unless necessary.

Let $\mathbf{y}_{\ell}^n$ denote the sequence of $\mathbf{y}_{\ell}$ from time slot~$1$ up to time slot~$n$, for 
$\ell \in \mathcal{U}$ and $n \in \mathcal{N}\!\!=\!\!\left\{1,2,\ldots,N\right\}$.
Now, we define the encoder and decoder functions for the considered system model \cite{IT_Cover}.
The encoder function at node~$i$ maps its own messages $W_{ij}$, $W_{ik}$ and $W_{i,\text{BC}}$, and the past values of the received symbols $\mathbf{y}_i^{n-1}$ into the symbol $\mathbf{x}_{i}\!\left(n\right)$. Therefore, the encoder function $\mathcal{E}_i$ of node $i$ is expressed as
\begin{IEEEeqnarray}{lCl} \label{eqn:encoding}
	\displaystyle
	\mathbf{x}_{i}\!\left(n\right) & = & \mathcal{E}_i \!\left(W_{ij}, W_{ik}, W_{i,\text{BC}}, \mathbf{y}_i^{n-1}\right),
\end{IEEEeqnarray}
where $i, j, k \in \mathcal{U} \text{ and } i \neq j \neq k$.
On the other hand, for a transmission block of length $N$, the decoder function at node~$i$ maps its own messages $W_{ij}$, $W_{ik}$ and $W_{i,\text{BC}}$, and the received symbols in each block $\mathbf{y}_i^N$ to form estimates of its desired messages $\hat{W}_{ji}$, $\hat{W}_{ki}$, $\hat{W}_{j,\text{BC}}$ and $\hat{W}_{k,\text{BC}}$.
Therefore, the decoder function $\mathcal{D}_i$ of node $i$ is expressed as
\begin{IEEEeqnarray}{lCl} \label{eqn:decoding}
	\displaystyle
	\left(\hat{W}_{ji}, \hat{W}_{ki}, \hat{W}_{j,\text{BC}}, \hat{W}_{k,\text{BC}}\right) & = & 
	\mathcal{D}_i \!\left(W_{ij}, W_{ik}, W_{i,\text{BC}}, \mathbf{y}_i^{N}\right)\!\!.
\end{IEEEeqnarray}

In this work, we use the total DoF as the key performance metric to characterize the capacity behavior in the high 
signal-to-noise ratio (SNR) regime \cite{Cadambe_Kuser_InterfAlgnm}. 
The DoF of a unicast message $W_{ij}$ with a rate $R_{ij}$ (as a function of the SNR) is designated as $d_{ij}$, 
for $i,j \in \mathcal{U} \text{ and } i \!\neq\! j$. It is characterized~as
\begin{IEEEeqnarray}{lCl} \label{eqn:uni_dof}
    \displaystyle
    d_{ij} 		& = & \lim_{\text{SNR} \rightarrow \infty} \frac{R_{ij}\!\left(\text{SNR}\right)}
    {\log \left(\text{SNR}\right)}.
\end{IEEEeqnarray}
Furthermore, the DoF of a broadcast message $W_{k,\text{BC}}$ with a rate $R_{k,\text{BC}}$ 
is designated as $d_{k,\text{BC}}$, for $k \in \mathcal{U}$. It is characterized~as
\begin{IEEEeqnarray}{lCl} \label{eqn:brcst_dof}
    \displaystyle
    d_{k,\text{BC}} & = & \lim_{\text{SNR} \rightarrow \infty} \frac{R_{k,\text{BC}}\!\left(\text{SNR}\right)}
    {\log \left(\text{SNR}\right)}.
\end{IEEEeqnarray}
The total DoF of the MIMO 3-way channel, $d_{\scriptscriptstyle\sum}$, is defined as
\begin{equation} \label{eqn:sum_dof}
    \displaystyle
    d_{\scriptscriptstyle\sum} = \sum_{i=1}^3 \sum_{j=1, j \neq i}^3 \!\!\! d_{ij}
    + 2 \sum_{k=1}^3 d_{k,\text{BC}}.
\end{equation}
It should be noted that the DoF of broadcast messages is weighted by two since any broadcast message 
is desired by two nodes in the network \cite{Broadcast_Salah}.
In other words, the weighting factor of the DoF of a message represents the
number of nodes that desires such a message and does not consider it as interference.
Since each unicast message is desired by one node, and it is treated as interference by the other node, the DoF of unicast 
messages is weighted by one.
On the contrary, each broadcast message is desired by two nodes and, hence, the DoF of broadcast messages is weighted by two.

\section{Case I: Unicast Messages Only} \label{sec:uni_only}
In this section, we characterize the asymmetric total DoF of the full-duplex MIMO 3-way channel when only unicast messages are exchanged among the nodes.
The following theorem presents the main result of this section.
\begin{theorem} \label{thrm1}
The optimal total DoF of the MIMO 3-way channel, with $M_1~\geq M_2~\geq M_3$, where each node sends a unicast message to each of the other two nodes, is given by
\begin{IEEEeqnarray}{lCl}
	\displaystyle
	d_{\scriptscriptstyle\sum} & = &
	\min \left\{
	M_1 \!+\! \frac{M_2 \!+\! M_3 \!-\! M_1}{3},\: M_2 \!+\! M_3 \right\}.
\end{IEEEeqnarray}
\end{theorem}
\begin{IEEEproof}
The converse proof of Theorem~\ref{thrm1} is presented in Section~\ref{Converse_Th1}, together with the optimal antenna allocation at each node that can achieve the maximum total DoF of the system. Finally, the achievability proof of Theorem~\ref{thrm1} is presented in Section~\ref{Achv_Th1}.

\subsection{\textbf{Converse Proof of Theorem~\ref{thrm1}}} \label{Converse_Th1}
The proof is divided into three parts. First, the cut-set bounds are provided.
Next, the genie-aided bounds are derived.
Finally, the optimal antenna allocation at each node is derived in order to maximize the total DoF given by the cut-set and genie-aided bounds.
Under the unicast communication scenario, the total DoF of the MIMO 3-way channel is characterized as
\begin{equation}
    \displaystyle
    d_{\scriptscriptstyle\sum} = d_{12} + d_{13} + d_{21} + d_{23} + d_{31} + d_{32}.
\end{equation}

\subsubsection{\textbf{Cut-set Bounds}} 
%
The derivation of cut-set bounds hinges on the cut-set theorem \cite{IT_Cover}.
Let $\mathcal{S}$ and $\mathcal{S}^c$ denote the set of source and destination nodes, respectively, where $\mathcal{S}^c$ is the complement of $\mathcal{S}$.
We start the proof by arguing that the cooperation of any two nodes among the three nodes does not degrade the DoF \cite{IT_Cover}. Taking this fact into consideration, we first consider the cut around $\mathcal{S} = \left\{1\right\}$ and $\mathcal{S}^c = \left\{2,3\right\}$. This leads to the following inequality
\begin{equation} \label{eqn:Cs_1_23}
	d_{12} \!+\! d_{13} \leq  
	\min \left\{M_{T_{1}},\: M_{R_{2}} \!\!+\! M_{R_{3}}\right\}.
\end{equation} 
Similarly, the following upper bounds can be obtained
\begin{IEEEeqnarray}{lCl}
	d_{21} \!+\! d_{23} & \leq &  
	\min \left\{M_{T_{2}},\: M_{R_{1}} \!\!+\! M_{R_{3}}\right\},
	\label{eqn:Cs_2_13}
	\\
	d_{31} \!+\! d_{32} & \leq &  
	\min \left\{M_{T_{3}},\: M_{R_{1}} \!\!+\! M_{R_{2}}\right\}.
	\label{eqn:Cs_3_12}
\end{IEEEeqnarray}
Adding \eqref{eqn:Cs_1_23}, \eqref{eqn:Cs_2_13} and \eqref{eqn:Cs_3_12}, we get
\begin{IEEEeqnarray}{lCl} \label{eqn:Cs_dsum_1}
	\displaystyle
    d_{\scriptscriptstyle\sum} & \leq &
    \min \left\{M_{T_{1}} \!\!+\!\! M_{T_{2}} \!\!+\!\! M_{T_{3}}, 
    M_{T_{1}} \!\!+\!\! M_{T_{2}} \!\!+\!\! M_{R_{1}} \!\!+\!\! M_{R_{2}},
    \right.\nonumber\\
     & &
	M_{T_{1}} \!\!+\!\! M_{T_{3}} \!\!+\!\! M_{R_{1}} \!\!+\!\! M_{R_{3}}, 
	M_{T_{2}} \!\!+\!\! M_{T_{3}} \!\!+\!\! M_{R_{2}} \!\!+\!\! M_{R_{3}},
	\nonumber\\	
	 & &
	M_{T_{1}} \!\!+\!\! 2 M_{R_{1}} \!\!+\!\! M_{R_{2}} \!\!+\!\! M_{R_{3}}, 
	M_{T_{2}} \!\!+\!\! M_{R_{1}} \!\!+\!\! 2 M_{R_{2}} \!\!+\!\! M_{R_{3}},
	\nonumber\\	
	 & &
	\left.
	M_{T_{3}} \!\!+\!\! M_{R_{1}} \!\!+\!\! M_{R_{2}} \!\!+\!\! 2 M_{R_{3}},
	2 \left(\!M_{R_{1}} \!\!+\!\! M_{R_{2}} \!\!+\!\! M_{R_{3}}\!\right)
	\right\}.
\end{IEEEeqnarray}
On the other hand, if we consider the cut around $\mathcal{S} = \left\{1,2\right\}$ and 
$\mathcal{S}^c = \left\{3\right\}$, we obtain
\begin{equation} \label{eqn:Cs_12_3}
	d_{13} \!+\! d_{23} \leq  
	\min \left\{M_{T_{1}} \!\!+\! M_{T_{2}}, \: M_{R_{3}}\right\}.
\end{equation}
Similarly, the following upper bounds can be obtained
\begin{IEEEeqnarray}{lCl}
	d_{21} \!+\! d_{31} & \leq &  
	\min \left\{M_{T_{2}} \!\!+\! M_{T_{3}}, \: M_{R_{1}}\right\},
	\label{eqn:Cs_23_1}
	\\
	d_{12} \!+\! d_{32} & \leq &  
	\min \left\{M_{T_{1}} \!\!+\! M_{T_{3}}, \: M_{R_{2}}\right\}.
	\label{eqn:Cs_13_2}
\end{IEEEeqnarray}
Adding \eqref{eqn:Cs_12_3}, \eqref{eqn:Cs_23_1} and \eqref{eqn:Cs_13_2}, we get
\begin{IEEEeqnarray}{lCl} \label{eqn:Cs_dsum_2}
	\displaystyle
    d_{\scriptscriptstyle\sum} & \leq &
    \min \left\{M_{R_{1}} \!\!+\!\! M_{R_{2}} \!\!+\!\! M_{R_{3}}, 
    M_{T_{1}} \!\!+\!\! M_{T_{2}} \!\!+\!\! M_{R_{1}} \!\!+\!\! M_{R_{2}},
    \right.\nonumber\\
     & &
	M_{T_{1}} \!\!+\!\! M_{T_{3}} \!\!+\!\! M_{R_{1}} \!\!+\!\! M_{R_{3}}, 
	M_{T_{2}} \!\!+\!\! M_{T_{3}} \!\!+\!\! M_{R_{2}} \!\!+\!\! M_{R_{3}},
	\nonumber\\	
	 & &
	M_{R_{1}} \!\!+\!\! 2 M_{T_{1}} \!\!+\!\! M_{T_{2}} \!\!+\!\! M_{T_{3}}, 
	M_{R_{2}} \!\!+\!\! M_{T_{1}} \!\!+\!\! 2 M_{T_{2}} \!\!+\!\! M_{T_{3}},
	\nonumber\\	
	 & &
	\left.
	M_{R_{3}} \!\!+\!\! M_{T_{1}} \!\!+\!\! M_{T_{2}} \!\!+\!\! 2 M_{T_{3}},
	2 \!\left(M_{T_{1}} \!\!+\!\! M_{T_{2}} \!\!+\!\! M_{T_{3}}\right)
	\right\}\!.
\end{IEEEeqnarray}
Combining \eqref{eqn:Cs_dsum_1} and \eqref{eqn:Cs_dsum_2}, and then simplifying the resulting expression, the cut-set upper bound on the total DoF of the MIMO 3-way channel with unicast messages is characterized~as
\begin{IEEEeqnarray}{lCl} \label{eqn:Cs_dsum}
	\displaystyle
    d_{\scriptscriptstyle\sum} & \leq & 
    \min \left\{M_{T_{2}} \!\!+\!\! M_{T_{3}} \!\!+\!\! M_{R_{2}} \!\!+\!\! M_{R_{3}}, \:
    M_{T_{1}} \!\!+\!\! M_{T_{2}} \!\!+\!\! M_{T_{3}}, \right.
    \nonumber\\
    & & 
    \left. \qquad\:
    M_{R_{1}} \!\!+\!\! M_{R_{2}} \!\!+\!\! M_{R_{3}}
    \right\}.
\end{IEEEeqnarray}

In cut-set bounds, it is assumed that the nodes on the same side of the cut are fully cooperating.
For instance, if we consider the cut around $\mathcal{S} = \left\{1\right\}$ and $\mathcal{S}^c = \left\{2,3\right\}$, we can imagine a genie that transfers $W_{23}$ to node~$3$ and $W_{32}$ to node~$2$.
That is why, the cut-set bounds are referred to as the two-sided genie-aided bounds \cite{Genie_Mokhtar}.
In order to establish tighter bounds on the total DoF, we resort to the one-sided genie-aided bounds \cite{Genie_Mokhtar,YChannel_Chabaan,3way_Channel_Chabaan} which we refer to as the genie-aided bounds in the sequel.

\subsubsection{\textbf{Genie-aided Bounds}}
The key idea of genie-aided bounds is that we assume the genie transfers the side-information from one node to another and not the other way around \cite{Genie_Mokhtar}. For example, in cut-set bounds, the genie transfers $W_{23}$ and $W_{32}$ to nodes~$3$~and~$2$, respectively.
However, in genie-aided bounds, we assume that the genie transfers either $W_{23}$ or $W_{32}$ and, hence, the other message is not known at its respective node a~priori.

We assume that every node can decode its desired unicast messages from the other nodes, according to the decoding function in \eqref{eqn:decoding}, with an arbitrarily small probability of error. For example, node~$1$ decodes $W_{21}$ and $W_{31}$ using its received signal, $\mathbf{y}_1^N$, and its unicast messages, $W_{12}$ and $W_{13}$, intended to node~$2$ and node~$3$, respectively.
Thus, node~$1$ knows $\mathbf{y}_1^N$, $W_{12}$, $W_{13}$, $W_{21}$ and $W_{31}$ after the decoding process.
Node~$1$ cannot decode more messages without being provided with additional side-information. In order to decode more messages, node~$1$ should be more knowledgeable than some other nodes.
Suppose we want node~$1$ to be able to decode $W_{32}$. Knowing $W_{21}$, we should provide node~$1$ with $W_{23}$ and $\mathbf{y}_2^N$ in order to decode $W_{32}$.
Assume that the genie transfers $W_{23}$ to node~$1$ as side-information. 
Then, what is left is to specifically know the additional side-information that is required to be transferred by the genie in order to generate $\mathbf{y}_2^N$. We will elaborate this as follows.
Having $W_{21}$ and $W_{23}$, node~$1$ can generate $\mathbf{x}_2 \!\left(1\right)$.
We then evaluate the following expression
\begin{IEEEeqnarray}{lCl}
	\displaystyle
	\mathbf{y}_1 \!\left(1\right) \!-\! \mathbf{H}_{21} \mathbf{x}_2 \!\left(1\right)
	& = & \mathbf{H}_{21} \mathbf{x}_2 \!\left(1\right) \!+\! \mathbf{H}_{31} \mathbf{x}_3 \!\left(1\right)
	  \!+\! \mathbf{z}_1 \!\left(1\right) \!-\! \mathbf{H}_{21} \mathbf{x}_2 \!\left(1\right)
	\nonumber\\
	& = & \mathbf{H}_{31} \mathbf{x}_3 \!\left(1\right) \!+\! \mathbf{z}_1 \!\left(1\right).
\end{IEEEeqnarray}
Next, we multiply the previous expression by $\mathbf{H}_{31}^{\dag}$ to~get
\begin{IEEEeqnarray}{l}
	\displaystyle
	\mathbf{H}_{31}^{\dag} \left(\mathbf{y}_1 \!\left(1\right) \!-\! \mathbf{H}_{21} \mathbf{x}_2 \!\left(1\right)\right)
	= \mathbf{x}_3 \!\left(1\right) + \mathbf{H}_{31}^{\dag} \mathbf{z}_1 \!\left(1\right).
\end{IEEEeqnarray}
It is worth mentioning that the left pseudo-inverse of $\mathbf{H}_{31}$ is guaranteed to exist almost surely if and only if $M_{R_1} \!\geq\! M_{T_3}$. Let us assume that this condition holds true for now and then we will later study the case when this condition is not satisfied.
Taking into consideration Eq. \eqref{eqn:rx_sig}, node~$1$ generates $\mathbf{y}_2 \!\left(1\right)$ as~follows.
\begin{IEEEeqnarray}{l} \label{eqn:noiseCorrect}
	\displaystyle
	\mathbf{H}_{32} \left(\mathbf{x}_3 \!\left(1\right) + \mathbf{H}_{31}^{\dag} \mathbf{z}_1 \!\left(1\right)\right)
	+ \mathbf{H}_{12} \mathbf{x}_1 \!\left(1\right)
	\nonumber\\
	= \left(\mathbf{H}_{12} \mathbf{x}_1 \!\left(1\right) \!+\! \mathbf{H}_{32} \mathbf{x}_3 \!\left(1\right) 
	  \!+\! \mathbf{z}_2 \!\left(1\right)\right)
	  + \left(\mathbf{H}_{32} \mathbf{H}_{31}^{\dag} \mathbf{z}_1 \!\left(1\right) 
	  \!-\! \mathbf{z}_2 \!\left(1\right)\right)
	\nonumber\\
	= \mathbf{y}_2 \!\left(1\right) + \mathbf{g}_{1,W_{23}} \!\left(1\right),
\end{IEEEeqnarray}
where $\mathbf{g}_{1,W_{23}} \!\left(1\right) = 
\mathbf{H}_{32} \mathbf{H}_{31}^{\dag} \mathbf{z}_1 \!\left(1\right)\!-\! \mathbf{z}_2 \!\left(1\right)$.
We can see from \eqref{eqn:noiseCorrect} that the side-information that node~$1$ requires is 
$\mathbf{g}_{1,W_{23}} \!\left(1\right)$ and, hence, node~$1$ can subtract it from 
$\mathbf{H}_{23} \left(\mathbf{x}_3 \!\left(1\right) \!+\! \mathbf{H}_{31}^{\dag} \mathbf{z}_1 \!\left(1\right)\right)
\!+\! \mathbf{H}_{21} \mathbf{x}_1 \!\left(1\right)$ to generate $\mathbf{y}_2 \!\left(1\right)$.
Having $\mathbf{y}_2 \!\left(1\right)$, $W_{21}$ and $W_{23}$, node~$1$ can generate $\mathbf{x}_2 \!\left(2\right)$, according to the encoding function in \eqref{eqn:encoding}.
Following the same line of thought explained above, node~$1$ can accordingly generate $\mathbf{y}_2 \!\left(2\right)$.
Node~$1$ reiterates this procedure until it completely generates $\mathbf{y}_2^N$.

To sum up, when the genie transfers $W_{23}$ as well as $\mathbf{g}_{1,W_{23}}^N$ to node~$1$ as side-information, it becomes more knowledgeable than node~$2$, that only has $W_{21}$, $W_{23}$ and $\mathbf{y}_2^N$. Hence, node~$1$ can decode $W_{32}$ in addition to $W_{21}$ and $W_{31}$.
From Fano's inequality, we can write
\begin{IEEEeqnarray}{l}
	\displaystyle
	N \left(R_{21} + R_{31} + R_{32}\right)
	\nonumber\\
	\leq \textit{I} \left(\!\displaystyle\underbrace{W_{21}, W_{31}, W_{32}}_{\text{\normalfont$W_1$}};\: 
	  \mathbf{y}_1^N, \displaystyle\underbrace{W_{12}, W_{13}, W_{23}}_{\text{\normalfont$W_2$}}, 
	  \mathbf{g}_{1,W_{23}}^N\!\!\right)
	+ N \epsilon_N
	\nonumber\\
	= \textit{I} \left(W_1;\: \mathbf{y}_1^N, W_2, \mathbf{g}_{1,W_{23}}^N\right) + N \epsilon_N
	\nonumber\\
	\stackrel{(a)}{=} \textit{I} \left(W_1;\: W_2, \mathbf{g}_{1,W_{23}}^N\right)
	  + \textit{I} \left(W_1;\: \mathbf{y}_1^N \:|\: W_2, \mathbf{g}_{1,W_{23}}^N\right)
	  + N \epsilon_N
	\nonumber\\
	\stackrel{(b)}{=} \textit{I} \left(W_1;\: \mathbf{y}_1^N \:|\: W_2, \mathbf{g}_{1,W_{23}}^N\right) + N \epsilon_N
	\nonumber\\
	= \textit{h} \left(\mathbf{y}_1^N \:|\: W_2, \mathbf{g}_{1,W_{23}}^N\right) 
	  - \textit{h} \left(\mathbf{y}_1^N \:|\: W_1, W_2, \mathbf{g}_{1,W_{23}}^N\right) + N \epsilon_N
	\nonumber\\
	\stackrel{(c)}{\leq} \textit{h} \left(\mathbf{y}_1^N\right) 
	  - \textit{h} \left(\mathbf{y}_1^N \:|\: W_1, W_2, \mathbf{g}_{1,W_{23}}^N\right) + N \epsilon_N
	\nonumber\\
	\stackrel{(d)}{=} \textit{h} \left(\mathbf{y}_1^N\right) 
	  - \displaystyle \sum_{n=1}^N 
	  \textit{h} \left(\mathbf{y}_1 \!\left(n\right) |\: \mathbf{y}_1^{n-1}, W_1, W_2, \mathbf{g}_{1,W_{23}}^N\right) 
	  + N \epsilon_N	
	\nonumber
	\end{IEEEeqnarray}
	\begin{IEEEeqnarray}{l}
	\leq \textit{h} \!\left(\mathbf{y}_1^N\right) 
	- \displaystyle \sum_{n=1}^N 
	  \textit{h} \!\left(\mathbf{y}_1 \!\left(n\right) |\: 
	  	\mathbf{y}_1^{n-1}\!, W_1, W_2, \mathbf{g}_{1,W_{23}}^N, \dotsc \right.
	\nonumber\\
	  \hspace{153pt} \left. \mathbf{y}_2^{n-1}\!, \mathbf{y}_3^{n-1}\right) + \: N \epsilon_N  
	\nonumber\\
	\stackrel{(e)}{=} \textit{h} \!\left(\mathbf{y}_1^N\right) 
	- \displaystyle \sum_{n=1}^N 
	  \textit{h} \!\left(\mathbf{y}_1 \!\left(n\right) |\: 
	  	\mathbf{y}_1^{n-1}\!, W_1, W_2, \mathbf{g}_{1,W_{23}}^N, \dotsc \right.
	\nonumber\\
	  \hspace{95pt} \left. \mathbf{y}_2^{n-1}, \mathbf{y}_3^{n-1}, \mathbf{x}_2^{n}, \mathbf{x}_3^{n}, 
	\mathbf{z}_1^{n-1}\right) 
		+ \: N \epsilon_N  
	\nonumber\\
	\stackrel{(f)}{=} \textit{h} \!\left(\mathbf{y}_1^N\right) 
	- \displaystyle \sum_{n=1}^N 
	  \textit{h} \!\left(\mathbf{z}_1 \!\left(n\right) |\: 
	  	\mathbf{y}_1^{n-1}\!, W_1, W_2, \mathbf{g}_{1,W_{23}}^N, \dotsc \right.
	\nonumber\\
	  \hspace{95pt} \left. \mathbf{y}_2^{n-1}, \mathbf{y}_3^{n-1}, \mathbf{x}_2^{n}, \mathbf{x}_3^{n}, 
	\mathbf{z}_1^{n-1}\right) 
		+ \: N \epsilon_N 
	\nonumber\\
	\stackrel{(g)}{=} \textit{h} \!\left(\mathbf{y}_1^N\right) 
	- \displaystyle \sum_{n=1}^N 
	  \!\textit{h} \!\left(\mathbf{z}_1 \!\left(n\right) |\: \mathbf{g}_{1,W_{23}}^N, \mathbf{z}_1^{n-1} \right) 
		+ N \epsilon_N  
	\nonumber\\
	= \:\textit{h} \!\left(\mathbf{y}_1^N\right) - \textit{h} \!\left(\mathbf{z}_1^{n} \:|\: \mathbf{g}_{1,W_{23}}^N\right) 
	  + N \epsilon_N
	\nonumber\\
	\leq \displaystyle \sum_{n=1}^N \!\textit{h}\! \left( \!\left[\:\mathbf{H}_{21} \mathbf{H}_{31}\right] \!
	  \left[ \begin{array}{c} \!\!\mathbf{x}_2 \left(n\right)\!\! \\ \!\!\mathbf{x}_3 \left(n\right)\!\! 
			 \end{array} \right] 
	  \!+ \mathbf{z}_1 \!\left(n\right)\!\right) + \mathcal{O} \!\left(1\right) + N \epsilon_N,
	  \nonumber\\
\end{IEEEeqnarray}
where $\mathcal{O} \!\left(1\right)$ is a term that is irrelevant to the DoF characterization, $\left(a\right)$~follows from the chain rule for mutual information, 
$\left(b\right)$~follows from the fact that $W_1$, $W_2$ and $\mathbf{g}_{1,W_{23}}^N$ are independent from each other and, hence, $\textit{I} \left(W_1;\: W_2, \mathbf{g}_{1,W_{23}}^N\right) = 0$,
$\left(c\right)$~follows from the fact that conditioning reduces entropy,
$\left(d\right)$~follows from the chain rule for entropy,
$\left(e\right)$~follows from the fact that $\mathbf{x}_i \!\left(n\right)$ is a function of $W_{ij}$, $W_{ik}$ and $\mathbf{y}_i^{n-1}$ for $i, j, k \in \mathcal{U} \text{ and } i \neq j \neq k$, and $\mathbf{z}_1 \!\left(n\right) \!=\! \mathbf{y}_1 \!\left(n\right) 
- \left(\mathbf{H}_{21} \: \mathbf{x}_{2} \!\left(n\right) + \mathbf{H}_{31} \: \mathbf{x}_{3} \!\left(n\right)\right)$, 
$\left(f\right)$~follows from the fact that 
$\textit{h} \left(\mathbf{H}_{21} \: \mathbf{x}_{2} \!\left(n\right) + \mathbf{H}_{31} \: \mathbf{x}_{3} \!\left(n\right) 	+ \mathbf{z}_1 \!\left(n\right) |\: \mathbf{x}_2 \!\left(n\right), \mathbf{x}_3 \!\left(3\right)\right) = 
 \textit{h} \left(\mathbf{z}_1 \!\left(n\right) |\: \mathbf{x}_2 \!\left(n\right), \mathbf{x}_3 \!\left(3\right)\right)$,
$\left(g\right)$~follows from the fact that
$\mathbf{z}_1 \!\left(n\right)$ and $\left\{\mathbf{y}_i^{n-1}, W_1, W_2, \mathbf{x}_j^{n} \right\}$ are independent, for 
$i \in \mathcal{U} \text{ and } j \in \mathcal{U} \backslash\! \left\{1\right\}$.
It should be noted that $\epsilon_N \rightarrow 0$ as $N \rightarrow \infty$.
Thus, when $M_{R_1} \!\geq\! M_{T_3}$, the total DoF of $W_{21}$, $W_{31}$ and $W_{32}$ is upper bounded~by
\begin{IEEEeqnarray}{lCl}
	\displaystyle
	N \left(d_{21} \!+\! d_{31} \!+\! d_{32}\right) & \leq &
	N \left(\operatorname{rank} \left(\left[\:\mathbf{H}_{21} \mathbf{H}_{31}\:\right]\right) + \epsilon_N\right)
	\nonumber\\
	& = &	
	N \left(\min \!\left\{M_{R_1}, M_{T_2} \!\!+\!\! M_{T_3}\right\} + \epsilon_N\right)\!.
	\nonumber\\
\end{IEEEeqnarray}
When dividing both sides by $N$ and then letting $N \rightarrow \infty$, we obtain
\begin{IEEEeqnarray}{lCl} \label{eqn:uni_DoF_U1_W1_C1}
	\displaystyle
	d_{21} \!+\! d_{31} \!+\! d_{32} & \leq &	
	\min \!\left\{M_{R_1}, M_{T_2} \!\!+\!\! M_{T_3}\right\}\!,
	\text{ if } M_{R_1} \!\geq\! M_{T_3}.
	\nonumber\\
\end{IEEEeqnarray}
On the other hand, when $M_{T_3} \!\geq\! M_{R_1}$, the left pseudo-inverse of $\mathbf{H}_{31}$ does not exist.
To tackle this problem, we deduce an upper bound on the total DoF by increasing the number of receive antennas at 
node~$1$ such that $M_{R_1} \!=\! M_{T_3}$.
As a result, the total DoF of $W_{21}$, $W_{31}$ and $W_{32}$ is upper bounded~by
\begin{IEEEeqnarray}{lCl} \label{eqn:uni_DoF_U1_W1_C2}
	\displaystyle
	d_{21} \!+\! d_{31} \!+\! d_{32} & \leq &
	\min \!\left\{M_{T_3}, M_{T_2} \!\!+\!\! M_{T_3}\right\}\!,
	\text{ if } M_{T_3} \!\geq\! M_{R_1}.
	\nonumber\\
\end{IEEEeqnarray}
Combining \eqref{eqn:uni_DoF_U1_W1_C1} and \eqref{eqn:uni_DoF_U1_W1_C2}, we finally get
\begin{equation} \label{eqn:uni_DoF_U1_W1}
	\displaystyle
	d_{21} \!+\! d_{31} \!+\! d_{32} \leq
	\min \left\{\max \left\{M_{R_1}, M_{T_3}\right\}\!, M_{T_2} \!+\! M_{T_3}\right\}.
\end{equation}

We have based our previous discussion on the assumption that the genie provides node~$1$ with $W_{23}$ and $\mathbf{g}_{1,W_{23}}^N$ to be able to decode $W_{32}$.
Now we assume that the genie transfers $W_{32}$ and $\mathbf{g}_{1,W_{32}}^N$ to node~$1$ in order to decode $W_{23}$.
Following the same approach, we can find that 
\begin{IEEEeqnarray}{lCl}
\mathbf{g}_{1,W_{32}}^N & = & \mathbf{H}_{23} \mathbf{H}_{21}^{\dag} \mathbf{z}_1^N - \mathbf{z}_3^N.
\end{IEEEeqnarray}
Therefore, the total DoF of $W_{21}$, $W_{31}$ and $W_{23}$ is upper bounded by
\begin{equation} \label{eqn:uni_DoF_U1_W2}
	\displaystyle
	d_{21} + d_{31} + d_{23} \leq
	\min \left\{\max \left\{M_{R_1}, M_{T_2}\right\}\!, M_{T_2} \!+\! M_{T_3}\right\}.
\end{equation}

Following the same procedure, we can derive the genie-aided bounds from node~$2$ and node~$3$ perspectives as follows.
For node~$2$, when the genie provides it with $W_{13}$ and 
$\mathbf{g}_{2,W_{13}}^N = \mathbf{H}_{31} \mathbf{H}_{32}^{\dag} \mathbf{z}_2^N - \mathbf{z}_1^N$,
the total DoF of $W_{12}$, $W_{13}$ and $W_{31}$ is upper bounded by
\begin{equation} \label{eqn:uni_DoF_U2_W1}
	\displaystyle
	d_{12} + d_{32} + d_{31} \leq
	\min \left\{\max \left\{M_{R_2}, M_{T_3}\right\}\!, M_{T_1} \!+\! M_{T_3}\right\}.
\end{equation}
On the other hand, when the genie provides node~$2$ with $W_{31}$ and 
$\mathbf{g}_{2,W_{31}}^N = \mathbf{H}_{13} \mathbf{H}_{12}^{\dag} \mathbf{z}_2^N - \mathbf{z}_3^N$,
the total DoF of $W_{12}$, $W_{13}$ and $W_{13}$ is upper bounded by
\begin{equation} \label{eqn:uni_DoF_U2_W2}
	\displaystyle
	d_{12} + d_{32} + d_{13} \leq
	\min \left\{\max \left\{M_{R_2}, M_{T_1}\right\}\!, M_{T_1} \!+\! M_{T_3}\right\}.
\end{equation}
For node~$3$, when the genie provides it with $W_{12}$ and 
$\mathbf{g}_{3,W_{12}}^N = \mathbf{H}_{21} \mathbf{H}_{23}^{\dag} \mathbf{z}_3^N - \mathbf{z}_1^N$,
the total DoF of $W_{12}$, $W_{13}$ and $W_{21}$ is upper bounded by
\begin{equation} \label{eqn:uni_DoF_U3_W1}
	\displaystyle
	d_{13} + d_{23} + d_{21} \leq
	\min \left\{\max \left\{M_{R_3}, M_{T_2}\right\}\!, M_{T_1} \!+\! M_{T_2}\right\}.
\end{equation}
On the other hand, when the genie provides node~$3$ with $W_{21}$ and 
$\mathbf{g}_{3,W_{21}}^N = \mathbf{H}_{12} \mathbf{H}_{13}^{\dag} \mathbf{z}_3^N - \mathbf{z}_2^N$,
the total DoF of $W_{12}$, $W_{13}$ and $W_{12}$ is upper bounded by
\begin{equation} \label{eqn:uni_DoF_U3_W2}
	\displaystyle
	d_{13} + d_{23} + d_{12} \leq
	\min \left\{\max \left\{M_{R_3}, M_{T_1}\right\}\!, M_{T_1} \!+\! M_{T_2}\right\}.
\end{equation}
Adding \eqref{eqn:uni_DoF_U2_W1} and \eqref{eqn:uni_DoF_U3_W1}, we obtain
\begin{IEEEeqnarray}{lcl} \label{eqn:uni_sumDoF_1}
	\displaystyle
    d_{\scriptscriptstyle\sum} & \leq  \min & \left\{ 
    2 M_{T_1} \!\!+\!\! M_{T_2} \!\!+\!\! M_{T_3}, 
    M_{T_1} \!\!+\!\! M_{T_3} \!\!+\! \max \!\left\{M_{R_3}, M_{T_2}\right\}\!,
    \right.
    \nonumber\\
    & &
    M_{T_1} \!\!+\!\! M_{T_2} \!+\! \max \!\left\{M_{R_2}, M_{T_3}\right\},
    \nonumber\\
    & &
    \left.
    \max \!\left\{M_{R_2}, M_{T_3}\right\} \!+\! \max \!\left\{M_{R_3}, M_{T_2}\right\}
    \right\}.
\end{IEEEeqnarray}
Adding \eqref{eqn:uni_DoF_U1_W2} and \eqref{eqn:uni_DoF_U2_W2}, we get
\begin{IEEEeqnarray}{lcl} \label{eqn:uni_sumDoF_2}
	\displaystyle
    d_{\scriptscriptstyle\sum} & \leq  \min & \left\{ 
    M_{T_1} \!\!+\!\! M_{T_2} \!\!+\!\! 2 M_{T_3}, 
    M_{T_1} \!\!+\!\! M_{T_3} \!\!+\! \max \!\left\{M_{R_1}, M_{T_2}\right\}\!,
    \right.
    \nonumber\\
    & &
    M_{T_2} \!\!+\!\! M_{T_3} \!+\! \max \!\left\{M_{R_2}, M_{T_1}\right\},
    \nonumber\\
    & &
    \left.
    \max \!\left\{M_{R_2}, M_{T_1}\right\} \!+\! \max \!\left\{M_{R_1}, M_{T_2}\right\}
    \right\}.
\end{IEEEeqnarray}
Adding \eqref{eqn:uni_DoF_U1_W1} and \eqref{eqn:uni_DoF_U3_W2}, we obtain
\begin{IEEEeqnarray}{lcl} \label{eqn:uni_sumDoF_3}
	\displaystyle
    d_{\scriptscriptstyle\sum} & \leq  \min & \left\{ 
    M_{T_1} \!\!+\!\! 2 M_{T_2} \!\!+\!\! M_{T_3}, 
    M_{T_1} \!\!+\!\! M_{T_2} \!\!+\! \max \!\left\{M_{R_1}, M_{T_3}\right\}\!,
    \right.
    \nonumber\\
    & &
    M_{T_2} \!\!+\!\! M_{T_3} \!+\! \max \!\left\{M_{R_3}, M_{T_1}\right\},
    \nonumber\\
    & &
    \left.
    \max \!\left\{M_{R_3}, M_{T_1}\right\} \!+\! \max \!\left\{M_{R_1}, M_{T_3}\right\}
    \right\}.
\end{IEEEeqnarray}
Combining \eqref{eqn:uni_sumDoF_1}, \eqref{eqn:uni_sumDoF_2} and \eqref{eqn:uni_sumDoF_3} with the cut-set bounds given by \eqref{eqn:Cs_dsum}, the total DoF of the MIMO 3-way channel with unicast messages is upper bounded by
\begin{IEEEeqnarray}{lCl} \label{eqn:uni_sumDoF}
	\displaystyle
    d_{\scriptscriptstyle\sum} & \leq  \min & \left\{ 
    M_{T_1} \!\!+\!\! M_{T_2} \!\!+\!\! M_{T_3}, \:
    M_{R_1} \!\!+\!\! M_{R_2} \!\!+\!\! M_{R_3},
    \right.
    \nonumber\\
    & &
    \max \!\left\{M_{R_2}, M_{T_3}\right\} \!+\! \max \!\left\{M_{R_3}, M_{T_2}\right\},
    \nonumber\\
    & &
    \max \!\left\{M_{R_2}, M_{T_1}\right\} \!+\! \max \!\left\{M_{R_1}, M_{T_2}\right\},
    \nonumber\\
    & &
    \left.
    \max \!\left\{M_{R_3}, M_{T_1}\right\} \!+\! \max \!\left\{M_{R_1}, M_{T_3}\right\}
    \right\}.
\end{IEEEeqnarray}
\begin{corollary} \label{cr1}
The special case of $M_{T_1} \!=\! M_{T_2} \!=\! M_{T_3} \!=\! M_T$ and $M_{R_1} \!=\! M_{R_2} \!=\! M_{R_3} \!=\! M_R\:$, studied by Maier~\emph{et~al.} in~\cite{3way_Channel_Chabaan}, is covered by \eqref{eqn:uni_sumDoF}. In this case, the total DoF of the symmetric MIMO 3-way channel is upper bounded by
\begin{IEEEeqnarray}{lCl}
	\displaystyle
	d_{\scriptscriptstyle\sum} & \leq &
	\left\{ \,
	\begin{IEEEeqnarraybox}[][c]{L?L}
	\IEEEstrut
	\min \left\{3 M_R, 2 M_T\right\}  & \mbox{for } M_T \!\geq\! M_R, \\
	\min \left\{3 M_T, 2 M_R\right\}  & \mbox{for } M_T \!\leq\! M_R.
	\IEEEstrut
	\end{IEEEeqnarraybox}
	\right.
\end{IEEEeqnarray}
\end{corollary}

\subsubsection{\textbf{Optimal Antenna Allocation}}
In this part, we seek the optimal allocation of transmit and receive antennas at each node in terms of $M_1$, $M_2$ and $M_3$ to maximize the upper bound on the total DoF of the MIMO 3-way channel with unicast messages, given by~\eqref{eqn:uni_sumDoF}.
The optimization problem is formulated as follows
\begin{IEEEeqnarray}{lCl} \label{P1}
    \textbf{P1: } \:\: & \max_{d_{\scriptscriptstyle\sum}, M_{T_{\ell}}, M_{R_{\ell}}} &  
    	d_{\scriptscriptstyle\sum}
    \nonumber\\ 
    & \text{s.t. } &   
    	\eqref{eqn:uni_sumDoF},
    \nonumber\\ 
    & &
    	M_{T_{\ell}} \!+\! M_{R_{\ell}} \!=\! M_{\ell}, \text{ for } \ell \in \left\{1, 2, 3\right\}\!\!.
\end{IEEEeqnarray}
\begin{lemma} \label{lemma1}
The total DoF of the MIMO 3-way channel with unicast messages only is upper bounded by
\begin{IEEEeqnarray}{lCl}
	\displaystyle
	d_{\scriptscriptstyle\sum} & \leq & d_{\scriptscriptstyle\sum}^{\star},
\end{IEEEeqnarray}
where $d_{\scriptscriptstyle\sum}^{\star}$ is the optimal solution of \textbf{P1}, which is given by
\begin{IEEEeqnarray}{lCl} \label{Eqn_Opt_DOF}
	\displaystyle
	d_{\scriptscriptstyle\sum}^{\star} & = &
	\left\{ \,
	\begin{IEEEeqnarraybox}[][c]{L?L}
	\IEEEstrut
	M_1 \!+\! \frac{M_2 \!+\! M_3 \!-\! M_1}{3} & \mbox{for } M_1 \!\leq\! M_2 \!+\! M_3 \\
	M_2 \!+\! M_3                           & \mbox{for } M_1 \!\geq\! M_2 \!+\! M_3.
	\IEEEstrut
	\end{IEEEeqnarraybox}
	\right.
\end{IEEEeqnarray}
When $M_1 \!\leq\! M_2 \!+\! M_3$, one optimal antenna allocation that achieves the corresponding maximum total DoF is
\begin{IEEEeqnarray}{l} \label{mat_uni_soln1}
	\displaystyle
	\left[ M_{R_1}^{\star}, M_{R_2}^{\star}, M_{R_3}^{\star} \right] \!=\! 
	\left[ 0, \frac{M_1 \!+\! 2 M_2 \!-\! M_3}{3}, 
		\frac{M_1 \!+\! 2 M_3 \!-\! M_2}{3} \right]\!.
	\nonumber\\
\end{IEEEeqnarray}
On the other hand, when $M_1 \!\geq\! M_2 \!+\! M_3$, one optimal antenna allocation that yields the maximum total DoF in this case is
\begin{IEEEeqnarray}{l} \label{mat_uni_soln2}
	\displaystyle
	\left[ M_{R_1}^{\star}, M_{R_2}^{\star}, M_{R_3}^{\star} \right] \!=\! 
	\left[ M_2 \!+\! M_3,\: 0,\: 0 \right].
\end{IEEEeqnarray}
Note that $M_{T_{\ell}}^{\star} = M_{\ell} - M_{R_{\ell}}^{\star}$ according to the second constraint of~\textbf{P1}.
\end{lemma}
\begin{IEEEproof}
The details of the solution of \textbf{P1} are reported in Appendix~\ref{App1}.
This completes the converse proof of Theorem~\ref{thrm1}.
\end{IEEEproof}
Fig.~\ref{fig:Total_DoF_Unicast} depicts the optimal total DoF of~\textbf{P1}. 
In the first region where~$\frac{M_1}{M_3} \!\leq\! \frac{M_2}{M_3} \!+\! 1$, the maximum total DoF that can be achieved
is~$\frac{1}{3} \!\left(\frac{2 M_1}{M_3} \!+\! \frac{M_2}{M_3} \!+\! 1 \right)$.
On the other hand, in the second region where~$\frac{M_1}{M_3} \!\geq\! \frac{M_2}{M_3} \!+\! 1$, 
the maximum total DoF that can be achieved is~$\frac{M_2}{M_3} \!+\! 1$.

\begin{figure}
\centering
\includegraphics[width=1\linewidth]{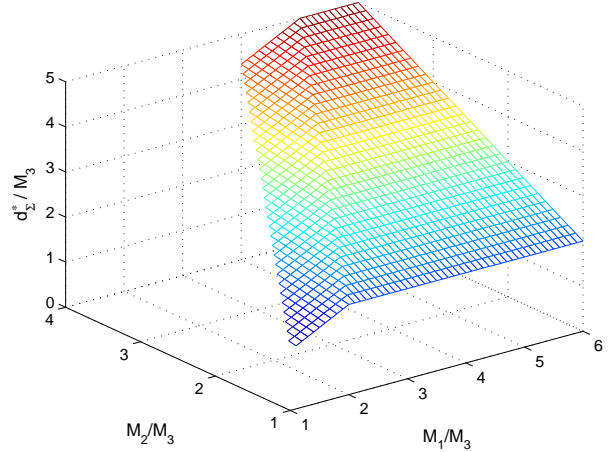}
\caption{The optimal total DoF of the full-duplex MIMO 3-way channel with unicast messages~only, 
for $M_1 \!\geq\! M_2 \!\geq\! M_3$.}
\label{fig:Total_DoF_Unicast}
\end{figure}

\subsection{\textbf{Achievability Proof of Theorem~\ref{thrm1}}} \label{Achv_Th1}
In this subsection, we provide the achievable schemes of total DoF of the MIMO 3-way channel described in 
Theorem~\ref{thrm1}.
Let $i, j, k \in \mathcal{U} \text{ and } i \neq j \neq k$.
A message $W_{ij}$ is encoded at the transmitter into the symbol $\mathbf{u}_{ij} \in \mathbb{C}^{r_{ij} \times 1}$, 
where $r_{ij} \leq M_{T_{i}}$.
The transmitted signal from node~$i$, $\mathbf{x}_{i} \in \mathbb{C}^{M_{T_{i}} \times 1}$, is defined as 
\begin{equation} \label{eqn:tx_sig}
	\displaystyle
    \mathbf{x}_i = 
    \mathbf{T}_{ij} \mathbf{u}_{ij} + 
    \mathbf{T}_{ik} \mathbf{u}_{ik},
\end{equation}
where $\mathbf{T}_{ij} \in \mathbb{C}^{M_{T_{i}} \times r_{ij}}$ is the precoding matrix for the signal transmitted from node~$i$ to node~$j$.

\subsubsection{\bm{$M_1 \leq M_2 + M_3$}}
In this case, the total DoF of the MIMO 3-way channel is bounded by 
$d_{\scriptscriptstyle\sum} \leq M_1 + \frac{M_2 + M_3 - M_1}{3}$.
The transmit and receive antennas at each node are allocated as follows
\begin{IEEEeqnarray}{lCllCl}
	\displaystyle
	M_{T_1} & = & M_1,\quad & 
	M_{R_1} & = & 0,
	\nonumber\\
	M_{T_2} & = & \frac{M_2 \!+\! M_3 \!-\! M_1}{3},\quad &
	M_{R_2} & = & \frac{M_1 \!+\! 2 M_2 \!-\! M_3}{3},
	\nonumber\\
	M_{T_3} & = & \frac{M_2 \!+\! M_3 \!-\! M_1}{3},\quad &
	M_{R_3} & = & \frac{M_1 \!+\! 2 M_3 \!-\! M_2}{3}.
\end{IEEEeqnarray}
It should be noted that if $M_{T_{\ell}}$ and $M_{R_{\ell}}$, for $\ell \in \mathcal{U}$, are not integers, we use the 
symbol extension method over multiple time slots~\cite{Jafar_Xchannel}. Then, we proceed with the design of the transmit 
strategy as explained below. 
Moreover, we assume in this achievable scheme that the number of antennas at each node is large enough to allow the allocation of transmit and receive antennas at the same time, i.e., $M_{\ell} \geq 3$ for $\ell \in \mathcal{U}$.
For example, if $M_2~=~1$, the proposed achievable scheme cannot be applied since this number of antennas cannot be partitioned, by any means, to allow simultaneous operation of the transmit and receive modes.
On the other hand, if $M_2 = 4$, we can apply the symbol extension method over three time~slots.
%
In the proposed scheme, all nodes transmit signals while nodes~$2$~and~$3$ receive signals.
Note that all antennas at node~$1$ are dedicated to signal transmission.
The~transmitted~signals~from each node are
\begin{IEEEeqnarray}{lCl}
	\displaystyle
	\mathbf{x}_1 & = & \mathbf{T}_{12} \mathbf{u}_{12} + \mathbf{T}_{13} \mathbf{u}_{13},
	\nonumber\\
	\mathbf{x}_2 & = & \mathbf{T}_{23} \mathbf{u}_{23},
	\nonumber\\
	\mathbf{x}_3 & = & \mathbf{T}_{32} \mathbf{u}_{32},
\end{IEEEeqnarray}
where the dimensions of encoded data symbols 
$\mathbf{u}_{12}$, $\mathbf{u}_{13}$, $\mathbf{u}_{23}$ and $\mathbf{u}_{32}$
are 
$\left(M_{T_1}\!\!-\!\!M_{R_3}\right) \!\times\! 1$, $\left(M_{T_1}\!\!-\!\!M_{R_2}\right) \!\times\! 1$, 
$M_{T_2} \!\times\! 1$ and $M_{T_3} \!\times\! 1$, respectively, 
whereas the dimensions of precoding matrices 
$\mathbf{T}_{12}$, $\mathbf{T}_{13}$, $\mathbf{T}_{23}$ and $\mathbf{T}_{32}$
are
$M_{T_1} \times \left(M_{T_1}\!\!-\!\!M_{R_3}\right)$, $M_{T_1} \times \left(M_{T_1}\!\!-\!\!M_{R_2}\right)$,
$M_{T_2} \!\times\! M_{T_2}$ and $M_{T_3} \!\times\! M_{T_3}$, respectively.
Note that $\mathbf{T}_{21} = \mathbf{T}_{31} = \mathbf{0}$ since $M_{R_1} = 0$.
The precoding matrices $\mathbf{T}_{12}$ and $\mathbf{T}_{13}$ are designed such that
\begin{IEEEeqnarray}{lCl}
	\displaystyle
	\mathbf{T}_{12} & \in & \operatorname{null} \left(\mathbf{H}_{13}\right),
	\nonumber\\
	\mathbf{T}_{13} & \in & \operatorname{null} \left(\mathbf{H}_{12}\right).
\end{IEEEeqnarray}
It is worth mentioning that the right pseudo-inverses of $\mathbf{H}_{13}$ and $\mathbf{H}_{12}$ exist almost surely owing to the fact that $M_{R_3} \!\leq\! M_{T_1}$ and $M_{R_2} \!\leq\! M_{T_1}$, respectively.
On the other hand, the precoding matrices $\mathbf{T}_{23}$ and $\mathbf{T}_{32}$ are randomly selected.
Consequently, the received signals at nodes~$2$~and~$3$ are
\begin{IEEEeqnarray} {lCl}
	\displaystyle
	\mathbf{y}_2 & = & 
	\mathbf{H}_{12} \mathbf{T}_{12} \mathbf{u}_{12} + \mathbf{H}_{32} \mathbf{T}_{32} \mathbf{u}_{32} + \mathbf{z}_2,
	\nonumber\\
	\mathbf{y}_3 & = & 
	\mathbf{H}_{13} \mathbf{T}_{13} \mathbf{u}_{13} + \mathbf{H}_{23} \mathbf{T}_{23} \mathbf{u}_{23} + \mathbf{z}_3.
\end{IEEEeqnarray}
Node~$2$ can decode $\mathbf{u}_{12}$ and $\mathbf{u}_{32}$ by projecting $\mathbf{y}_2$ to the null spaces of 
$\left(\mathbf{H}_{32} \mathbf{T}_{32}\right)^H$ and $\left(\mathbf{H}_{12} \mathbf{T}_{12}\right)^H$, respectively. 
Let $\mathbf{Q}_{12} \in \mathbb{C}^{M_{R_2} \times \left(M_{R_2} \!-\! M_{T_3}\right)}$ and 
$\mathbf{Q}_{32} \in \mathbb{C}^{M_{R_2} \times \left(M_{R_2} \!+\! M_{R_3} \!-\! M_{T_1}\right)}$ denote the projection matrices designed by node~$2$ such that
\begin{IEEEeqnarray}{lCl}
	\displaystyle
	\mathbf{Q}_{12} & \in & \operatorname{null} \left(\left(\mathbf{H}_{32} \mathbf{T}_{32}\right)^H\right),
	\nonumber\\	
	\mathbf{Q}_{32} & \in & \operatorname{null} \left(\left(\mathbf{H}_{12} \mathbf{T}_{12}\right)^H\right).
\end{IEEEeqnarray}
Since we assume that the nodes have prefect CSI knowledge, the zero-forcing estimates of $\mathbf{u}_{12}$ and $\mathbf{u}_{32}$ at node~$2$ are
\begin{IEEEeqnarray}{lCl}
	\displaystyle
	\hat{\mathbf{u}}_{12} & = & \mathbf{G}_{12} \left(\mathbf{Q}_{12}^H \mathbf{H}_{12} \mathbf{T}_{12}
	\mathbf{u}_{12} + \mathbf{Q}_{12}^H \mathbf{z}_2\right),
	\nonumber\\
	\hat{\mathbf{u}}_{32} & = & \mathbf{G}_{32} \left(\mathbf{Q}_{32}^H \mathbf{H}_{32} \mathbf{T}_{32}
	\mathbf{u}_{32}	+ \mathbf{Q}_{32}^H \mathbf{z}_3\right),
\end{IEEEeqnarray}
where $\mathbf{G}_{12} \in \mathbb{C}^{\left(M_{T_1}\!-\!M_{R_3}\right) \times \left(M_{T_1}\!-\!M_{R_3}\right)}$ and $\mathbf{G}_{32} \in \mathbb{C}^{M_{T_3} \times M_{T_3}}$ 
are the inverses of $\mathbf{Q}_{12}^H \mathbf{H}_{12} \mathbf{T}_{12}$ and 
$\mathbf{Q}_{32}^H \mathbf{H}_{32} \mathbf{T}_{32}$, respectively.
$\mathbf{G}_{12}$ and $\mathbf{G}_{32}$ are full rank almost surely because $\mathbf{Q}_{12}$ and $\mathbf{Q}_{32}$ are designed independently of $\mathbf{H}_{12}$ and $\mathbf{H}_{32}$, respectively, and $\mathbf{H}_{12}$ and $\mathbf{H}_{32}$ are drawn from a continuous random distribution.
Similarly, node~$3$ can decode $\mathbf{u}_{13}$ and $\mathbf{u}_{23}$.
As a result, node~$2$ decodes $M_{T_1} \!+\! M_{T_3} \!-\! M_{R_3}$ linearly independent information symbols while 
node~$3$ decodes $M_{T_1} \!+\! M_{T_2} \!-\! M_{R_2}$ linearly independent information symbols.
Thus, the scheme achieves a total of 
$2M_{T_1} \!\!+\!\! M_{T_2} \!\!+\!\! M_{T_3} \!\!-\!\! M_{R_2} \!\!-\!\! M_{R_3} \!=\!  
M_1 \!+\! \frac{M_2 + M_3 - M_1}{3}$ DoF for~$M_1 \!\leq\! M_2 \!+\! M_3$.

\subsubsection{\bm{$M_1 \geq M_2 + M_3$}}
In this case, the total DoF of the MIMO 3-way channel is bounded by 
$d_{\scriptscriptstyle\sum} \leq M_2 \!+\! M_3$.
The transmit and receive antennas at each node are allocated as follows
\begin{IEEEeqnarray}{lCllCl}
	\displaystyle
	M_{T_1} & = & M_1 \!-\! \left(M_2 \!+\! M_3\right),\quad & 
	M_{R_1} & = & M_2 \!+\! M_3,
	\nonumber\\
	M_{T_2} & = & M_2,\quad &
	M_{R_2} & = & 0,
	\nonumber\\
	M_{T_3} & = & M_3,\quad &
	M_{R_3} & = & 0.
\end{IEEEeqnarray}
%
In the proposed scheme, nodes~$2$~and~$3$ transmit signals to node~$1$.
The transmitted signals from nodes~$2$~and~$3$ are
\begin{IEEEeqnarray}{lCl}
	\displaystyle
	\mathbf{x}_2 & = & \mathbf{T}_{21} \mathbf{u}_{21},
	\nonumber\\
	\mathbf{x}_3 & = & \mathbf{T}_{31} \mathbf{u}_{31},
\end{IEEEeqnarray}
where $\mathbf{u}_{21} \in \mathbb{C}^{M_{T_2} \times 1}$ and $\mathbf{u}_{31} \in \mathbb{C}^{M_{T_3} \times 1}$,
whereas $\mathbf{T}_{21} \in \mathbb{C}^{M_{T_2} \times M_{T_2}}$ and 
$\mathbf{T}_{31} \in \mathbb{C}^{M_{T_3} \times M_{T_3}}$.
The precoding matrices $\mathbf{T}_{21}$ and $\mathbf{T}_{31}$ are randomly selected.
The received signal at node~$1$ is
\begin{IEEEeqnarray}{lCl}
	\displaystyle
	\mathbf{y}_1 & = & 
	\mathbf{H}_{21} \mathbf{T}_{21} \mathbf{u}_{21} + \mathbf{H}_{31} \mathbf{T}_{31} \mathbf{u}_{31} + \mathbf{z}_1.
\end{IEEEeqnarray}
Analogous to the previous case, node~$1$ applies zero-forcing to decode $\mathbf{u}_{21}$ and $\mathbf{u}_{31}$ separately. In other words, node~$1$ can decode $\mathbf{u}_{21}$ and $\mathbf{u}_{31}$ by designing $\mathbf{V}_{21}$ and $\mathbf{V}_{31}$ such that 
$\mathbf{V}_{21} \in \operatorname{null} \left(\left(\mathbf{H}_{31} \mathbf{T}_{31}\right)^H\right)$ and $\mathbf{V}_{31} \in \operatorname{null} \left(\left(\mathbf{H}_{21} \mathbf{T}_{21}\right)^H\right)$, respectively.
Afterwards, the zero-forcing estimates of $\mathbf{u}_{21}$ and $\mathbf{u}_{31}$ are obtained via evaluating the expressions $\mathbf{V}_{21}^H \mathbf{y}_1$ and $\mathbf{V}_{31}^H \mathbf{y}_1$, respectively.
As a result, node~$1$ decodes a total of $M_{T_2} \!+\! M_{T_3}$ independent information symbols are decoded and, hence, the scheme achieves $M_2 \!+\! M_3$ DoF for~$M_1 \!\geq\! M_2 \!+\! M_3$. 
This completes the achievability proof of Theorem~\ref{thrm1}.
\end{IEEEproof}

\section{Case II: Unicast and Broadcast Messages} \label{sec:uni_brcst}
In this section, we characterize the asymmetric total DoF of the full-duplex MIMO 3-way channel when unicast and broadcast messages are exchanged among the nodes.
The following theorem presents the main result of this section.
\begin{theorem} \label{thrm2}
The optimal total DoF of the MIMO 3-way channel, with $M_1~\geq M_2~\geq M_3$, where each node sends a unicast message to each of the other two nodes and a broadcast message to all other nodes, is given by
\begin{IEEEeqnarray}{lCl}
	\displaystyle
	d_{\scriptscriptstyle\sum} & = &
	M_2 \!+\! M_3.
\end{IEEEeqnarray}
\end{theorem}
\begin{IEEEproof}
The converse proof of Theorem~\ref{thrm2} is presented in Section~\ref{Converse_Th2}, together with the optimal antenna allocation at each node that can achieve the maximum total DoF of the system. Finally, the achievability proof of Theorem~\ref{thrm2} is presented in Section~\ref{Achv_Th2}.

\subsection{\textbf{Converse Proof of Theorem~\ref{thrm2}}} \label{Converse_Th2}
The proof is divided into two parts. First, the cut-set bounds are provided.
Second, the optimal antenna allocation at each node is derived in order to maximize the total DoF given by the cut-set bounds.
Under the unicast and broadcast communication scenario, the total DoF of the MIMO 3-way channel is characterized by \eqref{eqn:sum_dof}.

\subsubsection{\textbf{Cut-set Bounds}}
Let us consider the cut around $\mathcal{S}~=~\left\{1,2\right\}$ and 
$\mathcal{S}^c = \left\{3\right\}$. This leads to the following inequality
\begin{equation} \label{eqn:Cs_12_3_BC}
	d_{13} \!+\! d_{23} \!+\! d_{1,{\text{BC}}} \!+\! d_{2,{\text{BC}}} \leq  
	\min \left\{M_{T_{1}} \!\!+\! M_{T_{2}}, \: M_{R_{3}}\right\}.
\end{equation}
Similarly, the following upper bounds can be obtained
\begin{IEEEeqnarray}{lCl}
	d_{21} \!+\! d_{31} \!+\! d_{2,{\text{BC}}} \!+\! d_{3,{\text{BC}}} & \leq &  
	\min \left\{M_{T_{2}} \!\!+\! M_{T_{3}}, \: M_{R_{1}}\right\},
	\label{eqn:Cs_23_1_BC}
	\\
	d_{12} \!+\! d_{32} \!+\! d_{1,{\text{BC}}} \!+\! d_{3,{\text{BC}}} & \leq &  
	\min \left\{M_{T_{1}} \!\!+\! M_{T_{3}}, \: M_{R_{2}}\right\}.
	\label{eqn:Cs_13_2_BC}
\end{IEEEeqnarray}
Adding \eqref{eqn:Cs_12_3_BC}, \eqref{eqn:Cs_23_1_BC} and \eqref{eqn:Cs_13_2_BC}, and then simplifying the resulting expression,
the cut-set upper bound on the total DoF of the MIMO 3-way channel with unicast and broadcast messages is characterized~as
\begin{IEEEeqnarray}{lCl} \label{eqn:Cs_dsum_2_BC}
	\displaystyle
    d_{\scriptscriptstyle\sum} & \leq &
    \min \left\{M_{R_{1}} \!\!+\!\! M_{R_{2}} \!\!+\!\! M_{R_{3}}, 
    M_{T_{2}} \!\!+\!\! M_{T_{3}} \!\!+\!\! M_{R_{2}} \!\!+\!\! M_{R_{3}},
    \right.\nonumber\\
     & &
	\left.
	M_{R_{3}} \!\!+\!\! M_{T_{1}} \!\!+\!\! M_{T_{2}} \!\!+\!\! 2 M_{T_{3}},
	2 \!\left(M_{T_{1}} \!\!+\!\! M_{T_{2}} \!\!+\!\! M_{T_{3}}\right)
	\right\}\!.
\end{IEEEeqnarray}
After finding the optimal antenna allocation at each node that maximizes the total DoF of the system, it will be shown in the achievability proof that the cut-set bounds are tight and can be achieved due to the presence of broadcast messages.

\subsubsection{\textbf{Optimal Antenna Allocation}}
In this part, we seek the optimal allocation of transmit and receive antennas at each node in terms of $M_1$, $M_2$ and $M_3$ to maximize the upper bound on the total DoF of the MIMO 3-way channel with unicast and broadcast messages, given by~\eqref{eqn:Cs_dsum_2_BC}.
%
The optimization problem is formulated as follows
\begin{IEEEeqnarray}{lCl} \label{P2}
    \textbf{P2: } \: & \max_{d_{\scriptscriptstyle\sum}, M_{T_{\ell}}, M_{R_{\ell}}} &  
    	d_{\scriptscriptstyle\sum}
    \nonumber\\ 
    & \text{s.t. } &
    	\eqref{eqn:Cs_dsum_2_BC},  
    \nonumber\\ 
    & &
    	M_{T_{\ell}} \!+\! M_{R_{\ell}} \!=\! M_{\ell}, \text{ for } \ell \in \left\{1, 2, 3\right\}\!\!.
\end{IEEEeqnarray}
\begin{lemma} \label{lemma2}
The total DoF of the MIMO 3-way channel with unicast and broadcast messages is upper bounded by
\begin{IEEEeqnarray}{lCl}
	\displaystyle
	d_{\scriptscriptstyle\sum} & \leq & d_{\scriptscriptstyle\sum}^{\star},
\end{IEEEeqnarray}
where $d_{\scriptscriptstyle\sum}^{\star}$ is the optimal solution of \textbf{P2}, which is given by
\begin{IEEEeqnarray}{lCl} \label{brdcst_sumDoF}
	\displaystyle
	d_{\scriptscriptstyle\sum}^{\star} & = &
	M_2 \!+\! M_3.
\end{IEEEeqnarray}
Furthermore, one optimal antenna allocation that achieves the maximum total~DoF is
\begin{IEEEeqnarray}{lCl} \label{brdcst_OptSol_P2}
	\displaystyle
	\left[ M_{T_1}^{\star}, M_{T_2}^{\star}, M_{T_3}^{\star} \right] = 
	\left[ M_1 \!-\! M_2,\: M_2 \!-\! M_3,\: M_3 \right].
\end{IEEEeqnarray}
Note that $M_{R_{\ell}}^{\star} = M_{\ell} - M_{T_{\ell}}^{\star}$ according to the second constraint of~\textbf{P2}.
\end{lemma}
\begin{IEEEproof}
The details of the solution of \textbf{P2} are reported in Appendix~\ref{App2}.
This completes the converse proof of Theorem~\ref{thrm2}.
\end{IEEEproof}

\subsection{\textbf{Achievability Proof of Theorem~\ref{thrm2}}} \label{Achv_Th2}
In this subsection, we provide the achievable schemes of total DoF of the MIMO 3-way channel described in Theorem~\ref{thrm2}.
Let $i, j, k \in \mathcal{U} \text{ and } i \neq j \neq k$.
In addition to unicast messages, a broadcast message $W_{i,\text{BC}}$ is encoded at the transmitter into the symbol 
$\mathbf{u}_{i,\text{BC}} \in \mathbb{C}^{r_{i,\text{BC}} \times 1}$, where $r_{i,\text{BC}} \leq M_{T_{i}}$.
Accordingly, the transmitted signal from node~$i$, $\mathbf{x}_{i} \in \mathbb{C}^{M_{T_{i}} \times 1}$, is defined as 
\begin{equation}
	\displaystyle
    \mathbf{x}_i = 
    \mathbf{T}_{ij} \mathbf{u}_{ij} + 
    \mathbf{T}_{ik} \mathbf{u}_{ik} +
    \mathbf{T}_{i,\text{BC}} \mathbf{u}_{i,\text{BC}},
\end{equation}
where $\mathbf{T}_{i,\text{BC}} \in \mathbb{C}^{M_{T_{i}} \times r_{i,\text{BC}}}$ is the broadcast precoding matrix of node~$i$.

The total DoF of the MIMO 3-way channel is bounded by $d_{\scriptscriptstyle\sum} \leq M_2+M_3$.
The transmit and receive antennas at each node are allocated as follows
\begin{IEEEeqnarray}{lCllCl}
	\displaystyle
	M_{T_1} & = & M_1 \!-\! M_2,\quad & 
	M_{R_1} & = & M_2,
	\nonumber\\
	M_{T_2} & = & M_2 \!-\! M_3,\quad &
	M_{R_2} & = & M_3,
	\nonumber\\
	M_{T_3} & = & M_3,\quad &
	M_{R_3} & = & 0.
\end{IEEEeqnarray}
In the proposed scheme, nodes~$2$~and~$3$ transmit signals while nodes~$1$~and~$2$ receive signals.
Note that all antennas at node~$3$ are utilized for signal transmission.
The transmitted signals from nodes~$2$~and~$3$ are
\begin{IEEEeqnarray}{lCl}
	\displaystyle
	\mathbf{x}_2 & = & \mathbf{T}_{21} \mathbf{u}_{21},
	\nonumber\\
	\mathbf{x}_3 & = & \mathbf{T}_{3,\text{BC}} \mathbf{u}_{3,\text{BC}},
\end{IEEEeqnarray}
where $\mathbf{u}_{21} \in \mathbb{C}^{M_{T_2} \times 1}$ and 
$\mathbf{u}_{3,\text{BC}}  \in \mathbb{C}^{M_{T_3} \times 1}$, whereas
$\mathbf{T}_{21} \in~\mathbb{C}^{M_{T_2} \times M_{T_2}}$ and 
$\mathbf{T}_{3,\text{BC}} \in \mathbb{C}^{M_{T_3} \times M_{T_3}}$ which are selected randomly.
It is worth mentioning that $\mathbf{u}_{3,\text{BC}}$ is considered as a desired information symbol for  nodes~$1$~and~$2$. Therefore, it is not treated as interference by either node.
That is why, unlike unicast messages, broadcast messages provide additional degrees of freedom so that the cut-set bounds are tight and can be achieved.
The received signal at nodes~$1$~and~$2$~are 
\begin{IEEEeqnarray}{lCl}
	\displaystyle
	\mathbf{y}_1 & = & \mathbf{H}_{21} \mathbf{T}_{21} \mathbf{u}_{21} + 
					   \mathbf{H}_{31} \mathbf{T}_{3,\text{BC}} \mathbf{u}_{3,\text{BC}} + \mathbf{z}_1,
	\nonumber\\
	\mathbf{y}_2 & = & \mathbf{H}_{32} \mathbf{T}_{3,\text{BC}} \mathbf{u}_{3,\text{BC}} + \mathbf{z}_2.
\end{IEEEeqnarray}
Node~$2$ can decode $\mathbf{u}_{3,\text{BC}}$ using zero-forcing since $M_{R_2}~=~M_{T_3}$.
On the other hand, node~$1$ separates $\mathbf{u}_{21}$ from $\mathbf{u}_{3,\text{BC}}$ by designing the zero-forcing matrices $\mathbf{V}_{21}$ and $\mathbf{V}_{3,\text{BC}}$ of $M_{R_1} \!\times\! \left(M_{R_1}\!\!-\!\!M_{T_3}\right)$ and 
$M_{R_1} \!\times\! \left(M_{R_1}\!\!-\!\!M_{T_2}\right)$ dimensions, respectively, such that
\begin{IEEEeqnarray}{lCl}
	\displaystyle
	\mathbf{V}_{21} & \in & \operatorname{null} \left(\left(\mathbf{H}_{31} 
						  \mathbf{T}_{3,\text{BC}}\right)^H\right),
	\nonumber\\
	\mathbf{V}_{3,\text{BC}} & \in & \operatorname{null} \left(\left(\mathbf{H}_{21} 																				\mathbf{T}_{21}\right)^H\right).
\end{IEEEeqnarray}
Therefore, the following filtered signals are obtained
\begin{IEEEeqnarray}{lCl}
	\displaystyle
	\mathbf{V}_{21}^H \mathbf{y}_1 & = & 
	\mathbf{V}_{21}^H \mathbf{H}_{21} \mathbf{T}_{21} \mathbf{u}_{21}
	+ \mathbf{V}_{21}^H \mathbf{z}_1,
	\nonumber\\
	\mathbf{V}_{3,\text{BC}}^H \mathbf{y}_1 & = & 
	\mathbf{V}_{3,\text{BC}}^H \mathbf{H}_{31} \mathbf{T}_{3,\text{BC}} \mathbf{u}_{3,\text{BC}}
	+ \mathbf{V}_{3,\text{BC}}^H \mathbf{z}_1.
\end{IEEEeqnarray}
As a result, node~$1$ decodes $M_{T_2} \!+\! M_{T_3}$~linearly independent information symbols, and node~$2$ decodes $M_{T_3}$~linearly independent information symbols.
Thus, the scheme achieves a total of $M_{T_2} \!+\! 2 M_{T_3} \!=\! M_2 + M_3$~DoF 
for~$M_1 \!\geq\! M_2 \!\geq\! M_3$.
This completes the achievability proof of Theorem~\ref{thrm2}.
%
%
\end{IEEEproof}

\section{Conclusion} \label{sec:con}
In this paper, we characterized the total DoF of a MIMO 3-way channel with an asymmetric number of antennas at the nodes. 
Each node has a separate-antenna full-duplex MIMO transceiver where each antenna can be configured for either signal transmission or reception.
In the first message configuration, we considered the unicast communication scenario where each node can send two unicast messages to the two other nodes. 
The achievable total DoF is characterized using cut-set as well as genie-aided bounds. We rigorously derived the 
genie-aided bounds for the system in order to tighten the bounds given by the cut-set theorem.
In the second message configuration, we considered the unicast and broadcast communication scenario where each node can send two unicast messages as well as one broadcast message to the two other nodes. Due to the presence of broadcast messages, the cut-set bounds are tight and can be achieved. 
%
Next, we analytically derived the optimal allocation of transmit and receive antennas at each node in order to obtain the maximum total DoF for each message configuration, subject to the total number of antennas at each node.
Finally, we constructed the achievable schemes for each configuration using zero-forcing and null-space beamforming techniques.

\appendices
\section{} \label{App1}
In this appendix, we solve the optimization problem \textbf{P1}~in~\eqref{P1}.
This problem is non-convex due to the non-convexity of the feasible set.
In order to find the optimal solution of \textbf{P1}, we divide the non-convex feasible set into~$2^6$~polyhedrons, i.e., $2^6$~convex subsets, and then maximize the objective function over each subset.
The optimal solution of \textbf{P1} is obtained by selecting the solution that achieves the maximum value of the objective function among all the subproblems.
We can further reduce the number of subproblems to half, i.e.,~to~$2^5$~subproblems, by observing the symmetry of the objective function and the feasible set of \textbf{P1} in  
$\left[M_{T_1}, M_{T_2}, M_{T_3}\right]$ and $\left[M_{R_1}, M_{R_2}, M_{R_3}\right]$, which can be readily verified as follows.
Let $\left\{d_{\scriptscriptstyle\sum}^{\star}, M_{T_1}^{\star}, M_{T_2}^{\star}, M_{T_3}^{\star}, M_{R_1}^{\star}, M_{R_2}^{\star}, M_{R_3}^{\star}\right\}$ be the optimal solution of \textbf{P1}.
Substituting with $M_{T_1} \!=\! M_{R_1}^{\star}$, $M_{T_2} \!=\! M_{R_2}^{\star}$, $M_{T_3} \!=\! M_{R_3}^{\star}$, $M_{R_1} \!=\! M_{T_1}^{\star}$, $M_{R_2} \!=\! M_{T_2}^{\star}$ and $M_{R_3} \!=\! M_{T_3}^{\star}$ yields the same optimal value of the objective function while satisfying all the constraints of \textbf{P1}.
By~using the aforementioned approach, solving \textbf{P1} entails a one-dimensional line search with low complexity since the search space is finite and relatively small.
First, let us rewrite \textbf{P1} as follows
\begin{IEEEeqnarray}{lCl}
    \textbf{P3: } \:\: & \max_{d_{\scriptscriptstyle\sum}, M_{T_{\ell}}, M_{R_{\ell}}} \:\: &  
    	d_{\scriptscriptstyle\sum}
    \nonumber\\ 
    & \text{s.t. } &   
    	d_{\scriptscriptstyle\sum} \leq M_{T_1} \!\!+\!\! M_{T_2} \!\!+\!\! M_{T_3},
    \nonumber\\
    & &
    	d_{\scriptscriptstyle\sum} \leq M_{R_1} \!\!+\!\! M_{R_2} \!\!+\!\! M_{R_3},
    \nonumber\\
    & &
    	d_{\scriptscriptstyle\sum} \leq \max \!\left\{M_{R_2}, M_{T_3}\right\} \!+\! 
    	\max \!\left\{M_{R_3}, M_{T_2}\right\}\!\!,
    \nonumber\\
    & &
    	d_{\scriptscriptstyle\sum} \leq \max \!\left\{M_{R_2}, M_{T_1}\right\} \!+\! 
    	\max \!\left\{M_{R_1}, M_{T_2}\right\}\!\!,
    \nonumber\\
    & &
    	d_{\scriptscriptstyle\sum} \leq \max \!\left\{M_{R_3}, M_{T_1}\right\} \!+\! 
    	\max \!\left\{M_{R_1}, M_{T_3}\right\}\!\!,
    \nonumber\\
    & &
    	M_{T_{\ell}} \!+\! M_{R_{\ell}} \!=\! M_{\ell}, \text{ for } \ell \in \left\{1, 2, 3\right\}.
\end{IEEEeqnarray} 
It should be noted that the optimal solution of~\textbf{P1}~(and \textbf{P3}) relies on the values of $M_1$, $M_2$, and $M_3$, 
or more specifically, whether $M_1 \!\leq\! M_2 \!+\! M_3$ or $M_1 \!\geq\! M_2 \!+\! M_3$.
Therefore, we study each case separately.

\subsection{\bm{$M_1 \leq M_2 + M_3$}}
Let us consider one of the subproblems of \textbf{P3} and derive a closed-form expression of its optimal solution.
For instance, let us assume that $M_{R_2} \!\geq\! M_{T_3}$, $M_{R_3} \!\geq\! M_{T_2}$, $M_{T_1} \!\geq\! M_{R_2}$, 
$M_{T_2} \!\geq\! M_{R_1}$, $M_{T_1} \!\geq\! M_{R_3}$ and $M_{T_3} \!\geq\! M_{R_1}$.
Adding these assumptions to \textbf{P3} together with the constraint $M_1 \!\leq\! M_2 \!+\! M_3$, we get the following convex optimization problem
%
%
\begin{IEEEeqnarray}{lCl}
    \textbf{P4: } \:\: & \max_{d_{\scriptscriptstyle\sum}, M_{R_{\ell}}} \:\: &  
    	d_{\scriptscriptstyle\sum}
    \nonumber\\ 
    & \text{s.t. } &   
    	d_{\scriptscriptstyle\sum} \leq M_{R_2} \!+\! M_{R_3},
    \nonumber\\
    & &
    	d_{\scriptscriptstyle\sum} \leq M_1 \!+\! M_2 \!-\! M_{R_1} \!-\! M_{R_2},
    \nonumber\\
    & &
    	d_{\scriptscriptstyle\sum} \leq M_1 \!+\! M_3 \!-\! M_{R_1} \!-\! M_{R_3},
    \nonumber\\
    & &    	
    	d_{\scriptscriptstyle\sum} \geq 0, \:\:\: 0 \leq M_{R_{\ell}} \leq M_{\ell}, \text{ for } 
    	\ell \in \left\{1, 2, 3\right\},
   	\nonumber\\
    & &
    	M_{R_2} \geq M_{T_3}, \:\: M_{R_3} \geq M_{T_2},
    \nonumber\\
    & &
    	M_{T_1} \geq M_{R_2}, \:\: M_{T_2} \geq M_{R_1},
    \nonumber\\
    & &
    	M_{T_1} \geq M_{R_3}, \:\: M_{T_3} \geq M_{R_1},
    \nonumber\\
    & &
    	M_1 \leq M_2 \!+\! M_3.
\end{IEEEeqnarray}
\textbf{P4} can be expressed in the matrix form as follows
\begin{IEEEeqnarray}{lCl}
    \textbf{P5: } \:\: & \min_{\mathbf{v}} \:\: &  
    	\mathbf{c}^T \mathbf{v}
    \nonumber\\ 
    & \text{s.t. } &   
    	\mathbf{A} \mathbf{v} \preceq \mathbf{b},
\end{IEEEeqnarray}
where $\mathbf{v} = \left[d_{\scriptscriptstyle\sum}, M_{R_1}, M_{R_2}, M_{R_3}\right]^T$, 
$\mathbf{c} =\left[-1, 0, 0, 0\right]^T$,
\begin{equation}
\mathbf{A} = \left[
\begin{array}{cccc} 
\!\! 1  & 0  & -1  & -1  \!\!\\
\!\! 1  & 1  & 1  & 0  \!\!\\
\!\! 1  & 1  & 0  & 1  \!\!\\
\!\! 0  & 1  & 0  & 0  \!\!\\
\!\! 0  & 0  & 1  & 0  \!\!\\
\!\! 0  & 0  & 0  & 1  \!\!\\
\!\! -1  & 0  & 0  & 0  \!\!\\
\!\! 0  & -1 & 0  & 0  \!\!\\
\!\! 0  & 0  & -1 & 0  \!\!\\
\!\! 0  & 0  & 0  & -1 \!\!\\
\!\! 0  & 0  & -1 & -1  \!\!\\
\!\! 0  & 0  & -1 & -1  \!\!\\
\!\! 0  & 1 & 1 & 0  \!\!\\
\!\! 0  & 1 & 1 & 0  \!\!\\
\!\! 0  & 1 & 0 & 1 \!\!\\
\!\! 0  & 1 & 0 & 1 \!\!\\
\!\! 0  & 0 & 0 & 0 \!\!
\end{array}
\right]\! , \:\:
\mathbf{b} = \left[ 
\begin{array}{c} 
\!\! 0 \!\!\\ 
\!\! M_1 \!\!+\!\! M_2 \!\!\\ 
\!\! M_1 \!\!+\!\! M_3 \!\!\\
\!\! M_1 \!\!\\
\!\! M_2 \!\!\\
\!\! M_3 \!\!\\
\!\! 0 \!\!\\
\!\! 0 \!\!\\
\!\! 0 \!\!\\
\!\! 0 \!\!\\
\!\! - M_3 \!\!\\
\!\! - M_2 \!\!\\
\!\! M_1 \!\!\\
\!\! M_2 \!\!\\
\!\! M_1 \!\!\\
\!\! M_3 \!\!\\
\!\! M_2 \!\!+\!\! M_3 \!\!-\!\! M_1 \!\!
\end{array}
\right].
\end{equation}
It is obvious that \textbf{P5} is a linear program.
In order to find the optimal solution of \textbf{P5}, we establish the Lagrange dual problem of the primal problem as follows
\begin{IEEEeqnarray}{lCl}
    \textbf{P6: } \:\: & \max_{\boldsymbol\lambda} \:\: &  
    	-\mathbf{b}^T \boldsymbol\lambda
    \nonumber\\ 
    & \text{s.t. } &   
    	\mathbf{A}^T \boldsymbol\lambda + \mathbf{c} = 0,
    \nonumber\\
    & &
    	\boldsymbol\lambda \succeq 0,
\end{IEEEeqnarray}
where $\boldsymbol\lambda = \left[\lambda_1, \lambda_2, \ldots, \lambda_{16}\right]^T$.
If we can find a feasible point for \textbf{P5} and \textbf{P6} such that strong duality holds, i.e., the duality gap of the primal dual feasible pair, $\mathbf{c}^T \mathbf{v} + \mathbf{b}^T \boldsymbol\lambda$, is zero, then 
$\mathbf{v}^{\star}$~is primal optimal, $\boldsymbol\lambda^{\star}$ is dual optimal, and 
$d_{\scriptscriptstyle\sum}^{\star} = -\mathbf{c}^T \mathbf{v}^{\star} = \mathbf{b}^T \boldsymbol\lambda^{\star}$ \cite{ConvxOpt}.
%

Next, let~us assume that the $1^{\text{st}}$, $2^{\text{nd}}$, $3^{\text{rd}}$ and $8^{\text{th}}$ constraints of 
the primal problem~$\left(\textbf{P5}\right)$ are active, i.e., the inequality constraints are satisfied with equality.
Therefore, the primal problem reduces to a linear system of 4~equations and 4~unknowns and, hence, a feasible solution for the primal problem is obtained as
\begin{IEEEeqnarray}{lcl} \label{eqn:OptV}
	\displaystyle
	\mathbf{v} & = & 
	\left[
	\frac{2 M_1 \!+\! M_2 \!+\! M_3}{3},
	 0, \frac{M_1 \!+\! 2 M_2 \!-\! M_3}{3}, \frac{M_1 \!+\! 2 M_3 \!-\! M_2}{3} \right]^T\!\!\!\!.
	\nonumber\\
\end{IEEEeqnarray}
It is worth mentioning that this feasible solution yields
\begin{IEEEeqnarray}{lCl}
	\displaystyle
	\left[ M_{T_1}, M_{T_2}, M_{T_3} \right] \!=\! 
	\left[ M_1,\: \frac{M_2 \!+\! M_3 \!-\! M_1}{3},\: \frac{M_2 \!+\! M_3 \!-\! M_1}{3} \right].
	\nonumber\\
\end{IEEEeqnarray}
It is evident that the non-negativity constraint on $M_{T_2}$ and $M_{T_3}$ is satisfied only when 
$M_1 \!\leq\! M_2 \!+\! M_3$.
On the other hand, when strong duality holds, complementary slackness condition states that the $i$th optimal Lagrange multiplier $\lambda_i^{\star}$ is zero unless the $i$th inequality constraint is active at the optimum \cite{ConvxOpt}.
Taking this fact into consideration, let us assume that only the $1^{\text{st}}$, $2^{\text{nd}}$, $3^{\text{rd}}$ and $8^{\text{th}}$ elements of $\boldsymbol\lambda$ are non-zero.
Thus, the dual problem reduces to a linear system of 4~equations and 4~unknowns and, hence, a feasible solution for the dual problem is 
\begin{IEEEeqnarray}{lCl}
	\displaystyle
	\left[\lambda_1, \lambda_2, \lambda_3, \lambda_8\right] & = & 
	\left[\frac{1}{3},\: \frac{1}{3},\: \frac{1}{3},\: \frac{2}{3}\right].
\end{IEEEeqnarray}
Note that the resulting $\boldsymbol\lambda$ satisfies the non-negativity constraint.
For the obtained values of $\mathbf{v}$ and $\boldsymbol\lambda$, the corresponding duality~gap~is
\begin{IEEEeqnarray}{lCl}
	\displaystyle
	\mathbf{c}^T \mathbf{v} + \mathbf{b}^T \boldsymbol\lambda & = &  0.
\end{IEEEeqnarray}
Therefore, the primal dual feasible pair $\left(\mathbf{v}^{\star} \!, \boldsymbol\lambda^{\star}\right)$ is optimal and, hence, 
the maximum total DoF and the optimal antenna allocation of the considered subproblem are given by~\eqref{eqn:OptV}.

Similarly, we can formulate and solve all the remaining subproblems.
It turns out that the total DoF resulting from solving the aforementioned subproblem is the maximum value that can be attained from solving all the subproblems of \textbf{P1}. 
Moreover, due to the symmetry feature of \textbf{P1}, we can readily find the other optimal solution of \textbf{P1} that yields the same maximum total DoF and satisfies all the constraints.
Thus, when $M_1 \!\leq\! M_2 \!+\! M_3$, the optimal total DoF of~\textbf{P1} is given by~\eqref{Eqn_Opt_DOF}, and one optimal antenna allocation that achieves the maximum total DoF is given by~\eqref{mat_uni_soln1}.

\subsection{\bm{$M_1 \geq M_2 + M_3$}}
Following the same approach used in the previous case, we can find the optimal solution of~\textbf{P1} when 
$M_1 \!\geq\! M_2 \!+\! M_3$. It can be shown that the optimal total DoF of~\textbf{P1} is given by~\eqref{Eqn_Opt_DOF}, and one optimal antenna allocation that achieves the maximum total DoF is given by~\eqref{mat_uni_soln2}.
This completes the proof.

\section{} \label{App2}
In this appendix, we solve the optimization problem \textbf{P2}~in~\eqref{P2}.
To this end, we first rewrite \textbf{P2} as follows
\begin{IEEEeqnarray}{lCl}
    \textbf{P7: } \:\: & \max_{d_{\scriptscriptstyle\sum}, M_{T_{\ell}}} \:\: &  
    	d_{\scriptscriptstyle\sum}
    \nonumber\\ 
    & \text{s.t. } &   
    	d_{\scriptscriptstyle\sum} \leq M_1 \!+\! M_2 \!+\! M_3 \!-\! M_{T_1} \!\!-\!\! M_{T_2} \!\!-\!\! M_{T_3},
    \nonumber\\
    & &
    	d_{\scriptscriptstyle\sum} \leq M_2 \!+\! M_3,
    \nonumber\\
    & &
    	d_{\scriptscriptstyle\sum} \leq M_3 \!+\! M_{T_1} \!+\! M_{T_2} \!+\! M_{T_3},
    \nonumber\\
    & &
    	d_{\scriptscriptstyle\sum} \leq 2 M_{T_1} \!+\! 2 M_{T_2} \!+\! 2 M_{T_3},
    \nonumber\\
    & &    	
    	d_{\scriptscriptstyle\sum} \geq 0, \:\:\: 0 \leq M_{T_{\ell}} \leq M_{\ell}, \text{ for } 
    	\ell \in \left\{1, 2, 3\right\}.
    \nonumber\\
\end{IEEEeqnarray}
Next, let us consider the following optimization problem
\begin{IEEEeqnarray}{lCl}
    \textbf{P8: } \:\: & \max_{d_{\scriptscriptstyle\sum}, M_{T_{\ell}}} \:\: &  
    	d_{\scriptscriptstyle\sum}
    \nonumber\\ 
    & \text{s.t. } &   
    	0 \leq d_{\scriptscriptstyle\sum} \leq M_2 \!+\! M_3,
    \nonumber\\
    & &    	
    	\displaystyle \max \left\{d_{\scriptscriptstyle\sum} \!-\! M_3, \frac{d_{\scriptscriptstyle\sum}}{2}\right\} 
    		\leq \sum_{\ell=1}^3  \!M_{T_{\ell}} \leq \sum_{\ell=1}^3  \!M_{\ell} - d_{\scriptscriptstyle\sum},
    \nonumber\\
    & &    	
    	0 \leq M_{T_{\ell}} \leq M_{\ell}, \text{ for } \ell \in \left\{1, 2, 3\right\},
\end{IEEEeqnarray}
where the $2^{\text{nd}}$ constraint of~\textbf{P8} is the result of combining the~$1^{\text{st}}$, $3^{\text{rd}}$ and $4^{\text{th}}$ constraints of~\textbf{P7}.
It can be readily shown~that
\begin{IEEEeqnarray}{lCl} \label{eqn:Opt_Tau_P8}
	d_{\scriptscriptstyle\sum}^{\star} & = & M_2 \!+\! M_3.
\end{IEEEeqnarray}
Substituting \eqref{eqn:Opt_Tau_P8} in the $2^{\text{nd}}$ and $3^{\text{rd}}$ constraints of~\textbf{P8}, we get
\begin{IEEEeqnarray}{l} \label{eqn:Opt_Sol_P8}
	\displaystyle M_2 \:\leq\: M_{T_1}^{\star} \!+\! M_{T_2}^{\star} \!+\! M_{T_3}^{\star} \:\leq\: M_1.
	\nonumber\\
		0 \leq M_{T_{\ell}}^{\star} \leq M_{\ell}, \text{ for } \ell \in \left\{1, 2, 3\right\}.
\end{IEEEeqnarray}
Since \textbf{P2} and \textbf{P8} are equivalent optimization problems, the optimal total DoF of~\textbf{P2} is given by~\eqref{brdcst_sumDoF}.
Furthermore, according to~\eqref{eqn:Opt_Sol_P8}, there are many solutions of~\textbf{P2} that optimally allocate the transmit and receive antennas at each node to achieve the maximum total DoF.
One optimal antenna allocation, that satisfies the conditions in~\eqref{eqn:Opt_Sol_P8}, is given by~\eqref{brdcst_OptSol_P2}.
This completes the proof.

\begingroup
\newif\ifgobblecomma
\gobblecommafalse 
\edef\FZ{?}
\edef\KM{,}
\catcode`?=\active
\catcode`,=\active
\def?{\FZ\gobblecommatrue}
\def,{\ifgobblecomma\gobblecommafalse\else\KM\fi}
\bibliographystyle{IEEEbib}
\linespread{1}
\bibliography{myRef}

\begin{thebibliography}{10}

\bibitem{IBFD_Survey}
A.~Sabharwal, P.~Schniter, D.~Guo, D.~W. Bliss, S.~Rangarajan, and R.~Wichman,
\newblock ``In-band full-duplex wireless: {C}hallenges and opportunities,''
\newblock {\em IEEE Journal on Selected Areas in Communications}, vol. 32, no.
  9, pp. 1637--1652, Sept. 2014.

\bibitem{Rice_FullDuplex}
M.~Duarte and A.~Sabharwal,
\newblock ``Full-duplex wireless communications using off-the-shelf radios:
  {F}easibility and first results,''
\newblock {\em Asilomar Conference on Signals, Systems, and Computers}, pp.
  1558--1562, Nov. 2010.

\bibitem{MIDU}
E.~Aryafar, M.~A. Khojastepour, K.~Sundaresan, S.~Rangarajan, and M.~Chiang,
\newblock ``{MIDU}: {E}nabling {MIMO} full duplex,''
\newblock {\em Proceedings of the {ACM} International Conference on Mobile
  Computing and Networking ({M}obi{C}om)}, pp. 257--268, Aug. 2012.

\bibitem{Exp_FullDuplex}
M.~Duarte, C.~Dick, and A.~Sabharwal,
\newblock ``Experiment-driven characterization of full-duplex wireless
  systems,''
\newblock {\em IEEE Transactions on Wireless Communications}, vol. 11, no. 12,
  pp. 4296--4307, Dec. 2012.

\bibitem{FullDyplex_WiFi}
M.~Duarte, A.~Sabharwal, V.~Aggarwal, R.~Jana, K.~K. Ramakrishnan, C.~W. Rice,
  and N.~K. Shankaranarayanan,
\newblock ``Design and characterization of a full-duplex multiantenna system
  for {WiFi} networks,''
\newblock {\em IEEE Transactions on Vehicular Technology}, vol. 63, no. 3, pp.
  1160--1177, Mar. 2014.

\bibitem{Passive_SelfInterf_FD}
E.~Everett, A.~Sahai, and A.~Sabharwal,
\newblock ``Passive self-interference suppression for full-duplex
  infrastructure nodes,''
\newblock {\em IEEE Transactions on Wireless Communications}, vol. 13, no. 2,
  pp. 680--694, Feb. 2014.

\bibitem{SingAnt_FD}
M.~E. Knox,
\newblock ``Single antenna full duplex communications using a common carrier,''
\newblock {\em IEEE 13th Annual Wireless and Microwave Technology Conference
  (WAMICON)}, pp. 1--6, Apr. 2012.

\bibitem{Katti_fullDuplexRadios}
D.~Bharadia, E.~McMilin, and S.~Katti,
\newblock ``Full duplex radios,''
\newblock {\em Proceedings of the {ACM} {SIGCOMM}}, pp. 375--386, Aug. 2013.

\bibitem{5G_FullDuplex}
S.~Hong, J.~Brand, J.~I. Choi, M.~Jain, J.~Mehlman, S.~Katti, and P.~Levis,
\newblock ``Applications of self-interference cancellation in 5{G} and
  beyond,''
\newblock {\em IEEE Communications Magazine}, vol. 52, no. 2, pp. 114--121,
  Feb. 2014.

\bibitem{KumuNetworks}
``Kumu {N}etworks,''
\newblock www.kumunetworks.com.

\bibitem{Shannon_2way}
C.~E. Shannon,
\newblock ``Two-way communication channels,''
\newblock {\em 4th Berkeley Symposium on Mathematical Statistics and
  Probability}, vol. 1, pp. 611--644, 1961.

\bibitem{D2D_Chabban}
A.~Chaaban, H.~Maier, and A.~Sezgin,
\newblock ``The degrees-of-freedom of multi-way device-to-device communications
  is limited by 2,''
\newblock {\em IEEE International Symposium on Information Theory (ISIT)}, pp.
  361--365, Jul. 2014.

\bibitem{3way_CapReg_Chabban}
H.~Maier, A.~Chaaban, R.~Mathar, and A.~Sezgin,
\newblock ``Capacity region of the reciprocal deterministic 3-way channel via
  {$\Delta$}-{Y} transformation,''
\newblock {\em $52^{\text{nd}}$ Annual Allerton Conference on Communication,
  Control, and Computing (Allerton)}, pp. 167--174, Oct. 2014.

\bibitem{3way_Channel_Chabaan}
H.~Maier, A.~Chaaban, and R.~Mathar,
\newblock ``Symmetric degrees of freedom of the {MIMO} 3-way channel with
  ${M_{\text{T}}} \times {M_{\text{R}}}$ antennas,''
\newblock {\em IEEE Information Theory Workshop (ITW)}, pp. 92--96, Nov. 2014.

\bibitem{Det_YChannel_Chabaan}
A.~Chaaban and A.~Sezgin,
\newblock ``The capacity region of the linear shift deterministic
  {Y}-channel,''
\newblock {\em IEEE International Symposium on Information Theory (ISIT)}, pp.
  2457--2461, Jul. 2011.

\bibitem{Han_fullDuplx_2020}
S.~Han, C.-L. I, Z.~Xu, C.~Pan, and Z.~Pan,
\newblock ``Full duplex: Coming into reality in 2020?,''
\newblock {\em IEEE Global Communications Conference (GLOBECOM)}, pp.
  4776--4781, Dec. 2014.

\bibitem{IT_Cover}
T.~M. Cover and J.~A. Thomas,
\newblock {\em Elements of Information Theory},
\newblock Wiley-Interscience, 2006.

\bibitem{Cadambe_Kuser_InterfAlgnm}
V.~R. Cadambe and S.~A. Jafar,
\newblock ``Interference alignment and degrees of freedom of the {K}-user
  interference channel,''
\newblock {\em IEEE Transactions on Information Theory}, vol. 54, no. 8, pp.
  3425--3441, Aug. 2008.

\bibitem{Broadcast_Salah}
M.~Salah, A.~El-Keyi, M.~Nafie, and Y.~Mohasseb,
\newblock ``Achievable degrees of freedom on {K}-user {MIMO} multi-way relay
  channel with common and private messages,''
\newblock {\em Asilomar Conference on Signals, Systems, and Computers}, pp.
  973--977, Nov. 2015.

\bibitem{Genie_Mokhtar}
M.~Mokhtar, Y.~Mohasseb, M.~Nafie, and H.~El~Gamal,
\newblock ``On the deterministic multicast capacity of bidirectional relay
  networks,''
\newblock {\em IEEE Information Theory Workshop (ITW)}, pp. 1--5, Aug. 2010.

\bibitem{YChannel_Chabaan}
A.~Chaaban, K.~Ochs, and A.~Sezgin,
\newblock ``The degrees of freedom of the {MIMO} {Y}-channel,''
\newblock {\em IEEE International Symposium on Information Theory (ISIT)}, pp.
  1581--1585, Jul. 2013.

\bibitem{Jafar_Xchannel}
S.~A. Jafar and S.~Shamai~(Shitz),
\newblock ``Degrees of freedom region of the {MIMO} {X} channel,''
\newblock {\em IEEE Transactions on Information Theory}, vol. 54, no. 1, pp.
  151--170, Jan. 2008.

\bibitem{ConvxOpt}
S.~Boyd and L.~Vandenberghe,
\newblock {\em Convex Optimization},
\newblock Cambridge University Press, 2004.

\end{thebibliography}
\endgroup
\end{document}